%% file: aa61418-26.tex
\documentclass[]{aa} 
\usepackage{graphicx}
\usepackage{subcaption}
\usepackage{pgf}
\usepackage{txfonts}
\usepackage{longtable}
\usepackage{xspace}
\usepackage{xcolor}
\definecolor{myred}{rgb}{0.67571825, 0, 0.17578125}
\definecolor{myblue}{rgb}{0, 0.203125, 0.3828125}
\usepackage{hyperref}
\hypersetup{
    colorlinks=true,
    linkcolor=myblue,
    urlcolor=myblue,
    citecolor=myblue,
    filecolor=myblue
}
\usepackage[markup=underlined]{changes}
\usepackage{color}
\usepackage[normalem]{ulem}

\newcommand{\psj}{PSJ}

\newcommand\degdeg{$^\circ$\xspace}
\renewcommand\degr{$^\circ$\xspace}
\renewcommand\arcsec{$^{\prime\prime}$\xspace}

\newcommand\WN{2001~WN$_{5}$\xspace}

\newcommand{\hgg}{HG$_1$G$_2$\xspace}
\newcommand{\ggparams}{G$_1$G$_2$\xspace}

\begin{document} 
\title{
   Rotation, spectral type, and albedo of the potentially hazardous asteroid (153814)~2001~WN$_{\text 5}$
}
\subtitle{Prior to the 2028 June close approach}
\author{
     Jin Beniyama\inst{1,2} \corrauth{jbeniyama@oca.eu}, 
     Alexey V. Sergeyev\inst{1,3} \email{alexey.sergeyev@oca.eu},
     Konstantinos Odysseas Xenos\inst{1} \email{odysseas.xenos@oca.eu}, 
     Benoit Carry\inst{1} \email{benoit.carry@oca.eu},
     Petr Pravec\inst{4} \email{petr.pravec@asu.cas.cz}, 
     Petr Fatka\inst{4} \email{petr.fatka@asu.cas.cz}, 
     Peter Ku\v{s}nir\'ak\inst{4} \email{peter.kusnirak@asu.cas.cz},
     Larry Denneau\inst{5} \email{denneau@hawaii.edu},
     \and 
     Vasilij Shevchenko\inst{3} \email{v.g.shevchenko@karazin.ua}
}
\institute{
    Université Côte d'Azur, 
    Observatoire de la Côte d'Azur, CNRS, Laboratoire Lagrange, Bd de l'Observatoire, 
    CS 34229, 06304 Nice Cedex 4, France 
    \and
    Department of Earth and Planetary Science, The University of Tokyo, 7-3-1 Hongo, Bunkyo, Tokyo 113-0033, Japan
    \and
    Institute of Astronomy, V.N. Karazin Kharkiv National University, 35 Sumska Str., Kharkiv 61022, Ukraine
    \and
    Astronomical Institute ASCR, Fričova 298, Ondřejov 251 65, Czech Republic
    \and
    Institute for Astronomy, University of Hawaii, Honolulu, HI 96822, USA
}
\date{Received June 12, 2026 / Revised July 24, 2026}

\abstract
{
The potentially hazardous asteroid (153814)~2001~WN$_5$ will pass inside the lunar distance on June 26, 2028, offering a rare opportunity to characterize a kilometer-scale near-Earth asteroid at high angular resolution. 
However, previous taxonomic classifications have been inconsistent, and its physical properties remain insufficiently consolidated.
}
{
We aim to constrain the rotation state, shape, visible colors, geometric albedo, and taxonomy of 2001~WN$_5$ before its 2028 close approach.
}
{
We combined new photometry from the 1.54~m Danish Telescope (DK154) with archival and survey observations from the Transiting Exoplanet Survey Satellite (TESS),
Dark Energy Camera (DECam),
Zwicky Transient Facility (ZTF),
and the Asteroid Terrestrial-impact Last Alert System (ATLAS).
These data were used to refine the rotation period, investigate the spin-shape solution space, derive visible colors, and estimate the geometric albedo from phase curve slopes.
}
{
We measured a synodic rotation period of 
$4.263 \pm 0.002$~h and $4.259 \pm 0.009$~h from DK154 and TESS, respectively, consistent with previous lightcurve observations.
The available lightcurves do not uniquely constrain the sidereal rotation period, but the preferred pole solutions lie in the southern hemisphere
in ecliptic coordinates.
Visible colors from multiple independent datasets are consistent with the C-complex, while the TESS phase curve slopes give $p_{\rm V} = 0.13\pm0.04$, consistent with previous thermal-infrared albedo estimates.
Combining the visible colors, albedo, and published near-infrared spectra, we classify 2001~WN$_5$ as most likely a B-type asteroid.
The effective diameter is estimated to be $D = 0.81 \pm 0.13$~km using the $H$-$G$ model,
while the linear model yields a slightly smaller value of 
$0.74 \pm 0.11$~km.
Dynamical models indicate that \WN likely originated from the inner main belt via the $\nu_6$ resonance, with the Flora--Baptistina and Nysa--Polana complexes identified as the parent candidates.
}
{
During the 2028 encounter, 2001~WN$_5$ should reach an apparent diameter of about 0.5~arcsec, making it an excellent target for coordinated photometric, spectroscopic, and high-angular-resolution observations.
Observations during the 2026–2027 apparition will be essential for improving its spin and shape model before its 2028 close approach.
}
\keywords{
Minor planets, asteroids: general --
Minor planets, asteroids: individual: (153814)~2001 WN$_5$ --
Methods: observational --
Techniques: photometric
}
\titlerunning{Physical properties of (153814)~2001 WN$_5$}
\authorrunning{Beniyama et al.}
\maketitle
\nolinenumbers

\section{Introduction}
Close encounters of near-Earth asteroids (NEAs) with Earth are a concern for planetary defense and also provide exceptional opportunities for physical characterization.
Such encounters are accompanied by rapidly changing observing geometry and, in the most favorable cases, enable measurements that would otherwise be difficult, including direct or quasi-resolved imaging \citep{Merline2008, Merline2012, Reddy2022}, tighter constraints on spin states and shapes \citep[e.g.,][]{Kwiatkowski2021, Beniyama2022}, and tests of whether planetary tides can affect rotation states \citep{deLeon2013, Moskovitz2020, Benson2020, Ballouz2024}, as well as investigations of surface refreshing \citep{Binzel2010, Nesvorny2010, Kim2023}.

Among the NEAs approaching Earth in the second half of this decade, (153814)~2001~WN$_5$ is a particularly favorable target for various observations.
This potentially hazardous asteroid (PHA) of the Apollo type has an effective diameter of about 0.9~km \citep{Mainzer2011d, Masiero2021b, Myhrvold2022}, approximately three times larger than that of another PHA, (99942)~Apophis, which will also make a close approach to Earth in April 2029 \citep[e.g.,][]{Dotson2022}.
The synodic rotation period of \WN\ has been estimated from several visible lightcurves obtained in 2010 and 2019, yielding $4.253 \pm 0.001$~h and $4.254 \pm 0.018$~h in 2010 \citep{Skiff2019}, and $4.259 \pm 0.002$~h \citep{Warner2020} and $4.260 \pm 0.002$~h \citep{Monteiro2023} in 2019.
Using WISE W4 (22~$\mu$m) band data acquired over 30.2 days, \citet{Lam2023} reported a period of $4.1894 \pm 0.095$~h with a large amplitude of 1.0~mag.

The geometric albedo of \WN\ has been estimated to be $p_{\rm V} \sim 0.1$ from diameter constraints derived using thermal infrared observations \citep{Mainzer2011d, Masiero2021b, Myhrvold2022}.
Visible and near-infrared spectra of \WN\ have been included in several statistical studies of NEAs \citep{Thomas2014, Binzel2019, Marsset2022a}.
\citet{Thomas2014} obtained a 0.7–2.5~$\mu$m spectrum of \WN\ with the SpeX instrument at the NASA Infrared Telescope Facility (IRTF) on 2010 October 15, and classified it as a K-, L-, or Sq-type asteroid in the Bus–DeMeo taxonomy \citep{DeMeo2009}, assigning an \texttt{INDET} (indeterminate) label because the possible subclassifications span more than one taxonomic complex.
\citet{Binzel2019} classified \WN\ as an L-type asteroid based on their IRTF/SpeX spectrum obtained on 2010 October 12, three days prior to the observations of \citet{Thomas2014}, combined with an unpublished visible spectrum.
\citet{Marsset2022a} also classified \WN\ with their new IRTF/SpeX spectrum of \WN obtained in 2019 as a C- or X-type asteroid.
The radar circular polarization ratio (SC/OC) was estimated to be around 0.4 in observations conducted in 2010 and 2021 \citep{Taylor2021, Virkki2022}, which is consistent with both S- and C-type asteroids \citep[see Fig.~5 of][]{Zambrano-Marin2024}.

On June 26, 2028, \WN will pass inside the Moon’s orbit, at a geocentric distance of only $\sim0.65$ lunar distances (LD).
This represents a rare opportunity for an object of kilometer-scale size and offers a unique observational window.
However, no dedicated study has yet consolidated the available physical information on \WN.
Consequently, dedicated characterization is required to provide the accurate physical parameters necessary to exploit this rare opportunity.
In this paper, we present our lightcurve observations and analysis of archival TESS and DECam data, together with serendipitous observations from ZTF and ATLAS, to provide a consolidated view of \WN\ and to constrain its physical properties prior to its 2028 close approach.

\section{Data and methods\label{sec:data}}
\begin{figure}[ht]
\centering
\includegraphics[width=1.0\hsize]{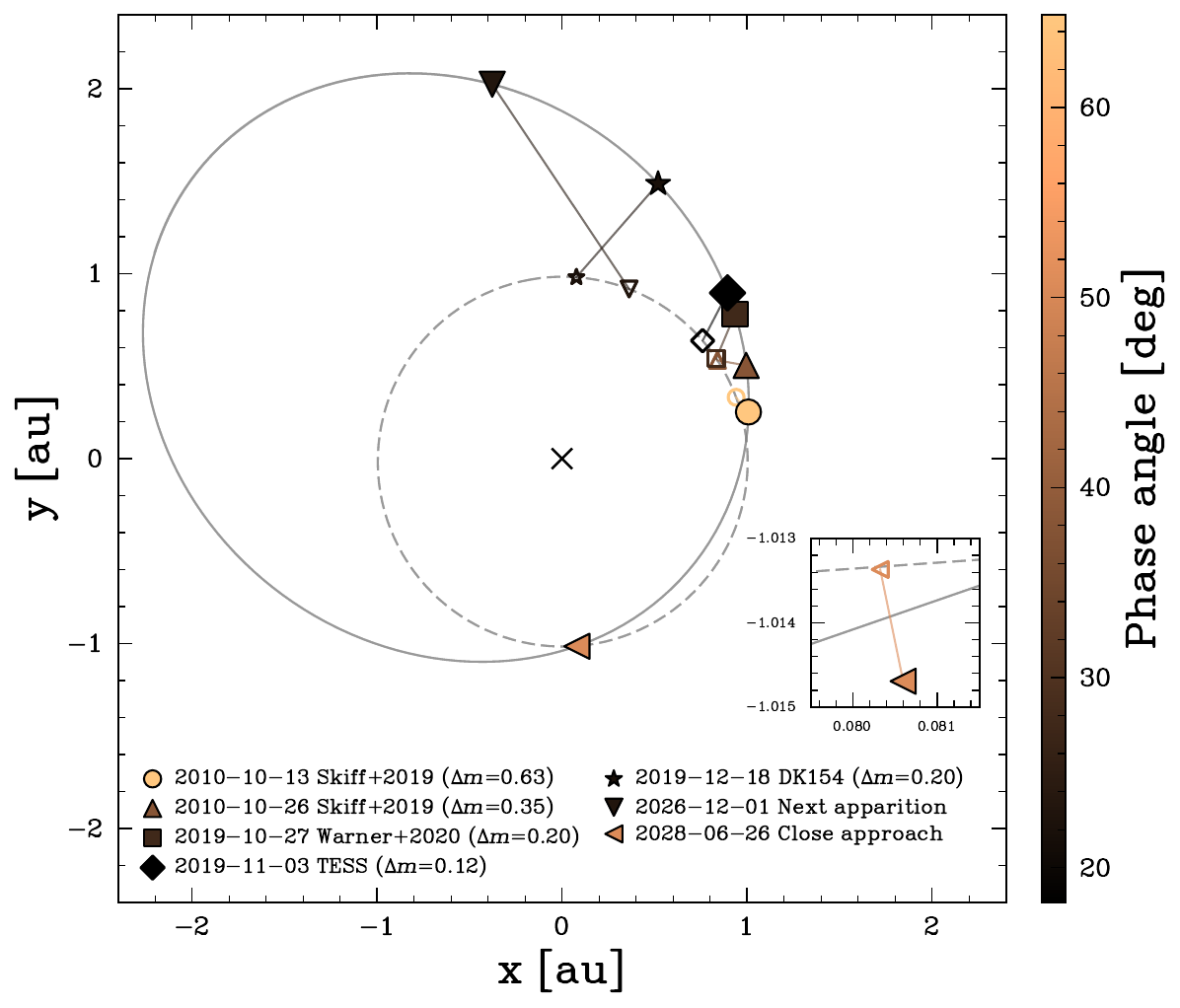}
\caption{
Heliocentric positions of \WN and Earth at the time of
lightcurve observations.
The positions of \WN and Earth are indicated by filled and open markers, respectively. 
Corresponding positions are connected by lines to clarify the observing geometry.
The orbits of \WN and Earth are shown with solid and dashed lines, respectively. 
The color of markers and solid lines indicates the solar phase angle of \WN.
The heliocentric positions of \WN at future favorable observing apparitions are also indicated.
The inset plot magnifies the specific region during the close approach in June 2028.
}
\label{fig:loc}
\end{figure}

We conducted lightcurve observations of \WN with the 1.54-m Danish Telescope (DK154),
and compiled publicly available data from the Transiting Exoplanet Survey Satellite (TESS),
Dark Energy Camera (DECam),
the Zwicky Transient Facility (ZTF), 
and 
the Asteroid Terrestrial-impact Last Alert System (ATLAS)
as well as historical measurements, spanning the period from 2010 to 2019.
The detailed observational circumstances, including the filter bands, coverage dates, predicted $V$ magnitudes, and phase angles, are summarized in Table~\ref{tab:obs_summary}.
These data cover a range of observing geometries, as shown in Fig.~\ref{fig:loc}.

\begin{table*}
\caption{Summary of the observational data used in this work.}
\label{tab:obs_summary}
\centering
\begin{tabular}{llccc}
\hline\hline
Source & Band & Date & $V$ & $\alpha$ \\
 & & (UTC)& (mag) & (deg) \\
\hline
\noalign{\smallskip}
DK154 & $V,R$ & 2019-12-18--2019-12-23 & 19.5--19.8 & 21.9--23.9 \\
TESS & TESS & 2019-11-04--2019-11-09 & 17.1--17.1 & 10.5--16.9 \\
DECam & $g, r$ & 2016-11-20--2016-11-22 & 22.7--22.7 & 15.5--16.0 \\
 & $g, i, z$ & 2019-02-14--2019-02-16 & 21.7--21.8 & 29.3--29.3 \\
ZTF & $g,r$ & 2019-11-03--2019-12-16 & 17.1--19.3 & 18.2--21.0 \\
ATLAS & $c,o$ & 2019-03-29--2019-12-28 & 17.0--20.0 & 3.4--100.0 \\
 \citet{Skiff2019} & $R$ & 2010-10-13--2010-10-26 & 15.8--16.1 & 38.1--64.9 \\
\citet{Warner2020} & $V$ & 2019-10-27--2019-10-30 & 17.0--17.1 & 23.6--27.8 \\
\noalign{\smallskip}\hline
\end{tabular}
\tablefoot{Predicted $V$-band apparent magnitude ($V$) range and solar phase angle ($\alpha$) range are referred to the NASA Jet Propulsion Laboratory (JPL) Horizons system for the corresponding observation epochs.}
\end{table*}

\subsection{The 1.54 m Danish Telescope (DK154)}
We observed \WN\ using DK154 and obtained time-series photometric data in the $V$ and $R$ bands on 5 nights from 18 to 23 December 2019.
The solar phase angle ranged from 22\degdeg to 24\degdeg.
The data were calibrated in the Johnson--Cousins photometric system using the Landolt standard stars \citep{Landolt1992}, with the nightly magnitude scale zero point uncertainties of $\pm 0.01$~mag. 
Since the observations were interleaved with other targets throughout each night, the resulting lightcurve is not a continuous time series but rather semi-dense lightcurves.
In total, 79~photometric data points were obtained.

On 3 of the 5 observing nights we took also series of 4+4 R and V images in quick succession, alternating between them (RVRVRVRV). 
From the data, we derived the 
rotation period and V-R color index by fitting the Fourier series of the 6th order 
(which is the maximum significant order of the DK154 lightcurve data, determined using the F-test; see \cite{Pravec2024}, Appendix B, and references therein) 
to the complete DK154 data while taking the difference between the V and R data as a free parameter.  
The estimated error of the determined color index ($\pm 0.02$) accounts for random errors as well as the absolute calibration errors of the data.

\subsection{Transiting Exoplanet Survey Satellite (TESS)}
We identified TESS observations of \WN\ acquired between 4 and 9 November 2019. The data were extracted as $11\times11$ pixel moving-target pixel files (TPFs) centred on the predicted position of \WN\ using the TESSCut service \citep{Brasseur2019}. The extracted images were corrected for background and then used to define a photometric aperture based on the point response function (PRF). Aperture photometry was subsequently performed using the \texttt{tess\_asteroids} package \citep{Tuson2026} to produce a lightcurve calibrated in TESS magnitudes.

The resulting lightcurve was cleaned by removing outliers with a sigma-clipping procedure. This yielded 254 useful measurements, each corresponding to a 30-min exposure, over a six-day observing interval. During this period, the solar phase angle of the asteroid ranged from 11\degr to 17\degr. These data were subsequently used to constrain both the rotation period and the phase curve of the asteroid (see Sects.~\ref{subsec:res_lc} and \ref{subsec:res_albedo}).
After correcting the magnitudes for the changing observing geometry, we determined the rotation period from the cleaned TESS lightcurve using an error-weighted generalized Lomb–Scargle periodogram \citep{Zechmeister2009}.

\subsection{Dark Energy Camera (DECam)}
We searched for serendipitous detections of \WN\ using the Solar System Object Image Search (SSOIS) at the Canadian Astronomy Data Centre \citep[CADC;][]{Gwyn2012}, using the object name \WN\ with default settings.
We identified several exposures from the Dark Energy Camera (DECam) on the 4-m Blanco telescope \citep{Flaugher2015}, although in some cases \WN\ lies outside the field of view.

In 2016, \WN\ was observed on 11 and 14 January in the $g$ and $r$ bands with exposure times ranging from 46 to 111 s. These observations were obtained as part of the DECam Legacy Survey of the SDSS Equatorial Sky \citep{Dey2019}, with the $g$ and $r$ exposures taken nearly simultaneously within two minutes.
In 2019, \WN\ was also serendipitously observed in multiple bands: $g$ on 14 February, $i$ on 27 February, and $z$ on 19 February, with exposure times of 70–90 s. These data were acquired as part of the DECam Local Volume Exploration Survey (\texttt{delve-wide}; \citealt{Drlica-Wagner2021}) and the DECam eROSITA Survey (\texttt{derositas}; \citealt{Zenteno2025}).

We downloaded and analyzed the calibrated, resampled images with filenames containing \texttt{opi}.
We performed circular aperture photometry of \WN. 
The aperture radii were optimised, and sufficiently large apertures were adopted to mitigate seeing variations across different bands.
Using the magnitude zero point recorded in the FITS header (\texttt{MAGZPT}), we computed magnitudes in the DES system.
We then converted the DES magnitudes to the SDSS system following the procedure of \citet{Carruba2024} and using the transformations of \citet{Abbott2021}.
For the $z$-band data, \WN was located near the edge of the detector chip, 
where significant background artifacts were observed. 
To account for these patterns, we instead employed rectangular aperture photometry for this specific band.
Cutout images are summarized in Appendix~\ref{app:cutout}.
We obtained seven photometric measurements: three in $g$,  two in $r$, and one each in $i$ and $z$.

\subsection{Zwicky Transient Facility (ZTF)}
We retrieved sparse photometry of \WN\ from ZTF \citep{Bellm2019} via the Fink
broker\footnote{\url{https://ztf.fink-portal.org/}} \citep{Moller2021}.
Specifically, we obtained 11 and 12 detections in the ZTF $g$ and $r$ bands, respectively.
The observations span the period from 2019 November 3 to December 16, corresponding to solar phase angles ranging from 3.4\degr to 21.1\degr in both bands.

\subsection{Asteroid Terrestrial-impact Last Alert System (ATLAS)}
We extracted sparse photometry of \WN\ from ATLAS \citep[][]{Tonry2018}, as recorded in the ATLAS Solar System Catalog (SSCAT) version 3.
The serendipitous detections are measured relative to the ATLAS Refcat photometric reference catalogue \citep{Tonry2018_catalog}.
We identified 55 and 250 detections in the cyan ($c$) and orange ($o$) bands, respectively, including observations at solar phase angles up to $100^{\circ}$.

Regarding the ATLAS data, an outlier rejection step (Xenos et al. in prep) preceded phase curve and shape modeling. Outliers in the data, associated with misidentifications of the asteroid with background sources, were flagged and removed using the offset between their observed and predicted astrometric positions and their reduced magnitude. This was followed by an iterative removal of the remaining outliers using the sHG$_1$G$_2$ phase curve model \citep{Carry2024}. 
As a result, 55 and 230 detections were retained in the 
$c$ and $o$ bands, respectively.

For phase-curve modeling, we selected a subset of 35 $c$-band detections from 2019 March 31 to November 28, and 126 $o$-band detections from 2019 March 29 to December 28, restricting the data to phase angles below 35~deg where the phase function is well constrained.
We note that the derived $c-o$ color remains consistent when the full ATLAS dataset is used.

\subsection{Historical measurements}
We have retrieved previously published lightcurves from the Asteroid Lightcurve Data Exchange Format (ALCDEF) database.
Specifically, two archival datasets were used.
The first set from \citet{Skiff2019}, obtained from LONEOS Schmidt (MPC code 699), consists of 825 measurements in the $SR$ band obtained on 2010 October 13, 14, and 26 (UTC). The corresponding solar phase angles decrease from 65\degr to 38\degr over these observing epochs.
The second dataset from \citet{Warner2020}, obtained 
at the Center for Solar System Studies (MPC code U81), comprises 289 measurements in the $V$ band and covers the period from 2019 October 27 to October 30 (UTC).
During this interval, the solar phase angles range from 28\degr to 24\degr.

\section{Results and discussion} \label{sec:result}
\subsection{Spin and shape} \label{subsec:res_lc}
\begin{figure}[ht]
\centering
\includegraphics[width=1.0\hsize]{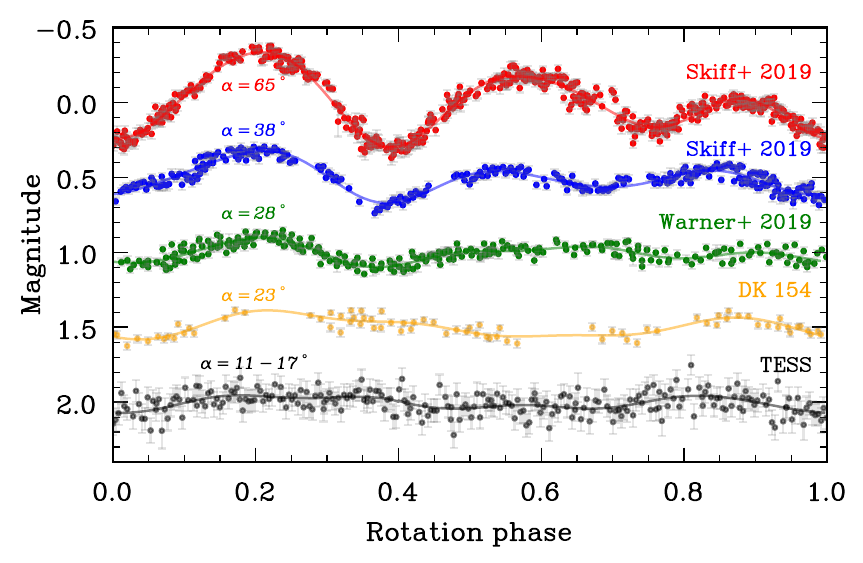}
\caption{
    Phased lightcurves of \WN obtained at different solar phase angles
    $\alpha$. The lightcurves were folded using the rotation period of $4.259$~h.
    The individual data sets are vertically shifted for clarity.
    }
\label{fig:lc}
\end{figure}

We determined the rotation periods of WN from the DK154 and TESS lightcurves to be $4.263 \pm 0.002$~h and $4.259 \pm 0.009$~h, respectively.
We showed the DK154 and TESS lightcurves in Fig.~\ref{fig:lc}, together with previously published lightcurves.

For each observing epoch, we measured the peak-to-peak amplitude from a smooth periodic fit to the phased lightcurve. 
The lightcurve amplitude ranged from 0.12 at a phase angle of $\sim14^\circ$ to 0.63 at a phase angle of $\sim65^\circ$.
The amplitudes decrease from $\Delta m = 0.63$ at a phase angle of $\sim65^\circ$ on 13 October 2010 to $\Delta m = 0.12$ at a mean phase angle of $\sim14^\circ$ on November 2019, 
with intermediate values of 
$\Delta m = 0.35$ at $\sim38^\circ$ on 26 October 2010, 
$\Delta m = 0.20$ at $\sim28^\circ$ on 27 October 2019,
and $\Delta m = 0.20$ at $\sim23^\circ$ in December 2019.
The presence of three minima and maxima within a single rotation indicates a complex lightcurve morphology, probably reflecting a non-ellipsoidal shape and/or surface-albedo variations. No clear evidence for non-principal-axis rotation is found in the available datasets.
This may be due to differences in solar phase angle; however, the possibility of changes in aspect angle cannot be ruled out \citep{Jackson2022, Carry2024}.

We performed shape modeling using the two methods:
lightcurve inversion method developed in \citet{Kaasalainen2001a} 
and the recent socca method, combining 
phase function with ellipsoidal shape \citep{xenos26}.
In the present case, this approach yielded multiple degenerate period solutions, 
preventing a unique determination of the sidereal period. 
Nevertheless, pole-orientation searches using the best-fit period favor a spin pole in the southern ecliptic hemisphere.

The details are summarized in Appendix~\ref{app:shape}.

\subsection{Albedo} \label{subsec:res_albedo}
\begin{figure}[ht]
\centering
\includegraphics[width=1.0\hsize]{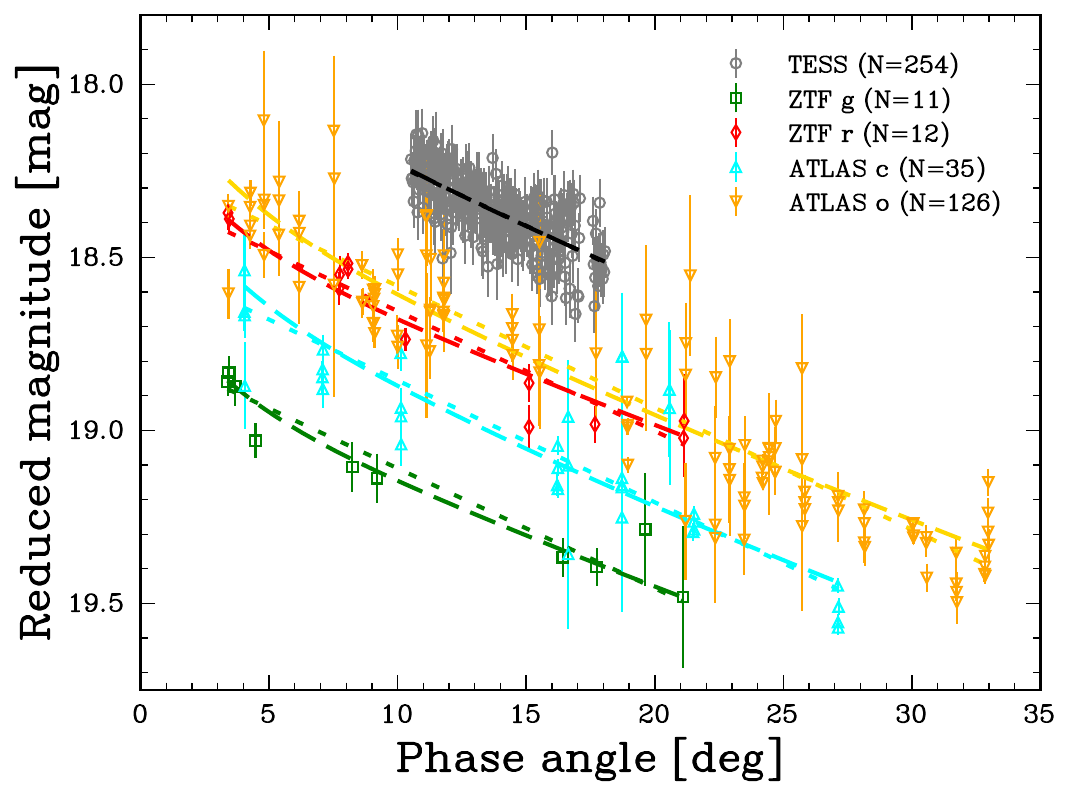}
\caption{
    Phase curves of \WN.
    The photometric data from different surveys and filters are represented by the following symbols: 
    TESS (circles), 
    ZTF $g$ (squares), ZTF $r$ (diamonds), ATLAS cyan (triangles), and ATLAS orange (inverted triangles).
    The dashed lines represent the best-fit curves using the $H$-$G$ model, while the dotted lines indicate the results from a linear model.
    }
\label{fig:pc}
\end{figure}

The extracted photometric data are shown as a function of solar phase angle in Fig. \ref{fig:pc}. The slope of a photometric phase curve serves as a diagnostic tool for determining albedo \citep{Belskaya2000, Shevchenko2021}. We estimated the slope using the $H$–$G$ model, defined as
\begin{equation}
H(\alpha) = H - 2.5 \log_{10}
\left[ (1 - G) \Phi_1(\alpha) + G \Phi_2(\alpha) \right],
\end{equation}
where $\Phi_1$ and $\Phi_2$ are the phase functions defined by \citet{Bowell1989}. 
Although the \hgg formalism \citep{2010Icar..209..542M} is more informative
on the surface properties, with the \ggparams correlated with
taxonomy and albedo
\citep[see, e.g.,][]{2016P&SS..123..101S},
it requires observations below 3--5\degr to provide 
meaningful parameters \citep{2021Icar..35414094M}.
As described in
Section~\ref{sec:data} and visible in Fig.~\ref{fig:pc}, 
there are no observations close to opposition, and we thus stick to the
$H$-$G$ formalism. 
We also applied a linear phase-curve model,
$H(\alpha)=H_{\rm linear}+b\alpha$,
because the presence of an opposition surge could not be established from the available data.

To minimize the influence of potential outliers in the photometric data, we derived the linear slope $b$ using the Siegel estimator, a robust regression method.
For the TESS data, we fitted the observations using both the $H$-$G$ system and a linear model, with $(H, G)$ and $(H_{\rm linear}, b)$ 
as the respective free parameters. 
For the ZTF and ATLAS data, we performed a simultaneous dual-band fit; specifically, we determined ($H_1, H_2, G$) for the $H$-$G$ model and ($H_1, H_2, b$) for the linear model.
The differences, $H_1-H_2$, are treated as colors (see the next subsection).
The parameter uncertainties were estimated using a Monte Carlo technique. 
We generated 10,000 synthetic phase curves by randomly resampling the dataset; in this process, each observed data point was assumed to follow a normal distribution, with the standard deviation defined by the standard error of the error-weighted average magnitude.

The derived parameters are summarized in Table \ref{tab:pc}. 
Using the empirical relation \citep{Belskaya2000, Shevchenko2021},
$b = C_1 - C_2 \log_{10}{p_{\rm V}}$,
where $C_1$ and $C_2$ are constants calibrated using asteroids with albedos determined from thermal infrared observations, we derive the geometric albedo from the slope. Adopting the updated parameters $C_1 = 0.016 \pm 0.001$ and $C_2 = 0.022 \pm 0.001$ \citep{Shevchenko2021}, we obtain geometric albedos for \WN of $p_{\rm V} = 0.13 \pm 0.04$, 
$0.10 \pm 0.04$,
and 
$0.14 \pm 0.02$ from 
the TESS, ZTF, and ATLAS photometry, respectively.

These values are in good agreement with those derived from thermal observations: $p_{\rm V}=0.097\pm0.016$ \citep{Mainzer2011d}, $p_{\rm V}=0.106^{+0.105}_{-0.053}$ \citep{Masiero2021b}, and $p_{\rm V}=0.089\pm0.021$ \citep{Myhrvold2022}. 
While the geometric albedo alone could also be consistent with an M-type asteroid, 
the available spectroscopic constraints make this interpretation unlikely (see the next subsection).
An albedo of $\sim0.1$ is somewhat higher than the typical value for Bus--DeMeo or Mahlke C-types and may instead be more consistent with B-types \citep{Marsset2022a}, supporting the interpretation that \WN\ is a relatively high-albedo member within the C-complex.

\begin{table*}
\caption{Best-fit parameters for the HG and Linear models.}
\label{tab:pc}
\centering
\begin{tabular}{llccccc}
\hline\hline
Instrument & Model & $H_1$ & $H_2$ & $H_1 - H_2$ & $G$ & $b$ \\
 & & (mag) & (mag) & (mag) & & (mag deg$^{-1}$) \\
\hline
\noalign{\smallskip}
TESS           & HG     & 17.56 $\pm$ 0.05 & -- & -- & 0.14 $\pm$ 0.04 & -- \\
               & Linear & 17.87 $\pm$ 0.04 & -- & -- & -- & 0.036 $\pm$ 0.003 \\
\noalign{\smallskip}\hline
ZTF ($g, r$)   & HG     & 18.55 $\pm$ 0.04 & 18.08 $\pm$ 0.05 & 0.47 $\pm$ 0.06 & 0.23 $\pm$ 0.07 & -- \\
               & Linear & 18.74 $\pm$ 0.03 & 18.27 $\pm$ 0.04 & 0.47 $\pm$ 0.05 & -- & 0.038 $\pm$ 0.003 \\
\noalign{\smallskip}\hline\noalign{\smallskip}
ATLAS ($c, o$) & HG     & 18.20 $\pm$ 0.03 & 17.94 $\pm$ 0.03 & 0.26 $\pm$ 0.04 & 0.14 $\pm$ 0.02 & -- \\
               & Linear & 18.54 $\pm$ 0.02 & 18.26 $\pm$ 0.02 & 0.28 $\pm$ 0.03 & -- & 0.035 $\pm$ 0.001 \\
\noalign{\smallskip}\hline\noalign{\smallskip}
\end{tabular}
\tablefoot{$H_1$ and $H_2$ denote the absolute magnitudes corresponding to the primary and secondary filters listed in the first column for each instrument (i.e., $H_1 = H_g$ and $H_2 = H_r$ for ZTF; $H_1 = H_c$ and $H_2 = H_o$ for ATLAS). For TESS, $H_1$ represents its single band. Parameters $G$ and $b$ are shared for ZTF and ATLAS dual-band fits.}
\end{table*}

\subsection{Colors and reflectance} \label{subsec:res_taxonomy}
The derived colors are summarized in Table \ref{tab:col}.
We computed the $g-z$ and $g-i$ colors of \WN using the $g$, $i$, and $z$ observations obtained in February 2019. 
The colors involving the DECam $z$-band, for instance $g-z$, appear anomalously blue, most likely because the observations were not obtained at the same rotation phase and/or because of the lower signal-to-noise ratio of the $z$-band detection. 
We therefore exclude the $z$-band measurement from the subsequent taxonomic interpretation.

\begin{figure}[ht]
\centering
\includegraphics[width=1.0\hsize]{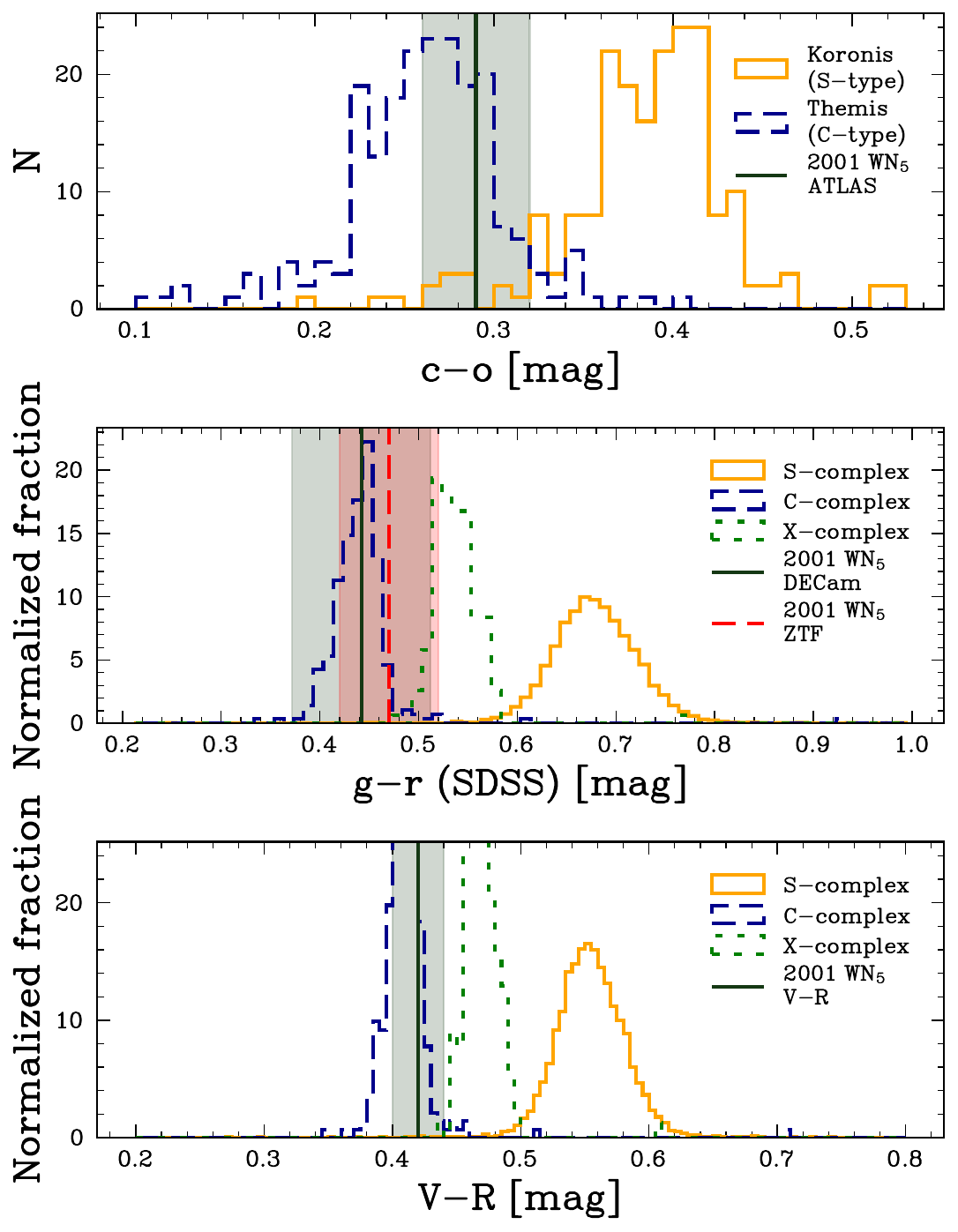}
\caption{
    Colors of \WN compared with reference asteroids. 
    (top) ATLAS $c-o$ color.
    The observed color of \WN is indicated by the vertical solid line.
    The distributions of the S-type Koronis family (solid line) and the C-type Themis family (dashed line) are shown for comparison \citep{Erasmus2020}.
    (middle) SDSS $g-r$ color. 
    The colors of \WN derived from different instruments are shown as vertical solid (DECam) and vertical dashed (ZTF) lines, respectively.
    The background distributions correspond to the S-complex (solid line), C-complex (dashed line), and X-complex (dotted line) \citep{Sergeyev2021}.
    (bottom)
    $V-R$ color.
    The observed color of \WN is indicated by the vertical solid line, plotted against the same taxonomic complex background distributions as in the middle panel.
}
\label{fig:col}
\end{figure}

In the top panel of Fig. \ref{fig:col}, 
we show the ATLAS colors of \WN together with those of two asteroid families from \citet{Erasmus2020}. 
Although the ATLAS bandpasses are broader than those of the SDSS system, the $c-o$ color index has been shown to be diagnostic of asteroid spectral type \citep{Erasmus2020}. The colors of \WN are consistent with a C-type classification.
In the middle panel of Fig. \ref{fig:col}, we present the SDSS colors of \WN together with asteroids observed in the SDSS survey \citep{Sergeyev2021}. For this comparison, we assume that the $g-r$ color in the ZTF system is equivalent to that in the SDSS system \citep{Schemel2021}. 
The colors derived from both DECam and ZTF are consistent with a C-complex classification.
In the bottom panel of Fig. \ref{fig:col}, we present the V-R colors of \WN together with asteroids observed in the SDSS survey \citep{Sergeyev2021}.

\begin{figure}[ht]
\centering
\includegraphics[width=1.0\hsize]{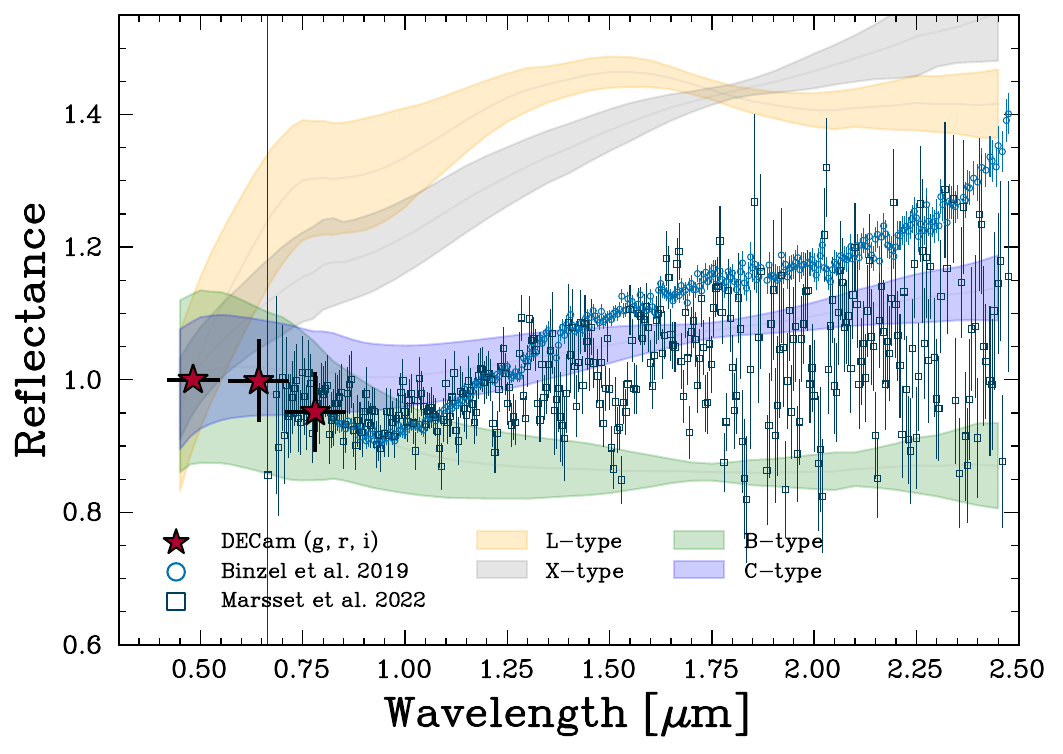}
\caption{
    Visible to near-infrared reflectance spectra of \WN normalized at the $g$ band center.
    The stars represent the reflectance values derived from our DECam photometry. 
    Previously published spectra are plotted with circles \citep[][]{Binzel2019} and squares \citep[][]{Marsset2022a}.
    The spectrum from \citet{Binzel2019} has been renormalized to match the DECam $i$-band reflectance, whereas the spectrum from \citet{Marsset2022a} remains normalized at 1.215~$\mu$m as originally published.
    Mahlke templates of L-, X-, B-, and C-types are shown \citep{Mahlke2022}. 
    Shaded areas indicate the uncertainties of the template spectra.
    }

\label{fig:spec}
\end{figure}

Overall, four independent color measurements, 
ATLAS, DECam, ZTF, and DK154, consistently indicate that \WN belongs to the C-complex 
\citep[e.g.,][]{2010A&A...510A..43C, 2013Icar..226..723D, 2021A&A...652A..59S, 2023A&A...679A.148S}.
In Fig. \ref{fig:spec}, we present the reflectance calculated from the derived colors and solar colors \citep{Willmer2018}, as well as existing spectra from the literature for
comparison\footnote{Several
near-infrared measurements of \WN from the Visible and Infrared Survey Telescope for Astronomy (VISTA) were obtained on July 13 and 24, 2010.
These were reported in the MOVIS catalog \citep{Popescu2016} as well as in the Minor Planet Center database.
Upon careful inspection, we found that the derived colors are inconsistent between the two epochs and appear unrealistic.
Therefore, we do not use these measurements in the analysis.
}.
The combined spectrum is consistent with a C-complex classification, and not with L-type or K-type, in the Mahlke taxonomy \citep{Mahlke2022}. 
Furthermore, when considering the albedo, the combined data are consistent with a bright C-complex classification, specifically a B-type.

    \begin{table*}
    \caption{Colors of \WN.}
    \label{tab:col}
    \centering
    \begin{tabular}{lccccc}
    \hline\hline
    Instrument & Color & Value & $\Delta t$ & Mid-time & Note \\
           &       & (mag) & (min)      & (UTC)    &      \\
    \hline
    DK154 & $V-R$ & 0.42 $\pm$ 0.02 & -- & -- & From lightcurves \\
DECam & $g-r$ & 0.41 $\pm$ 0.11 & 1.2 & 2016-01-11T06:39:31 &  \\
 & $g-r$ & 0.46 $\pm$ 0.08 & 1.3 & 2016-01-14T06:26:13 &  \\
 & $g-i$ & 0.48 $\pm$ 0.06 & 18758 & 2019-02-20T20:49:50 & Distance corrected \\
 & $g-z$ & 0.13 $\pm$ 0.10 & 7143 & 2019-02-16T20:02:14 & Distance corrected \\
ZTF & $g-r$ & 0.47 $\pm$ 0.05 & -- & -- & From phase curve \\
ATLAS & $c-o$ & 0.28 $\pm$ 0.03 & -- & -- & From phase curve \\
    \hline
    \end{tabular}
    \tablefoot{
    For DECam measurements, colors are computed from pairs of exposures.
    Mid-time is defined as the midpoint between the two filter observations.
    }
    \end{table*}

\subsection{Dynamical origin} \label{subsec:origin}
To investigate the origin of \WN, we referred to the near-Earth object source region probabilities provided by the Lowell AstOrb Database \citep{Moskovitz2022}, which are computed using the model of \citet{Granvik2018}. 
The orbital elements of \WN listed in the AstOrb Database are $a = 1.712$ au, $e = 0.467$, and $i = 1.920^\circ$, with its absolute magnitude of $H=18.280$.
The corresponding source region probabilities indicate a dominant contribution from the $\nu_6$ resonance (90.2\%).
This result suggests that \WN most likely originated from the inner main belt, with the $\nu_6$ resonance acting as the primary delivery pathway into near-Earth space.

We further investigated its possible parent asteroid family using the recent family-association model of \citet{Broz2024}\footnote{\url{https://sirrah.troja.mff.cuni.cz/~mira/neomod/neomod.php}}. 
Based on the orbital elements of \WN, the family-association model assigns the highest probabilities to the Flora (34.5\%) and Polana (28.4\%) families, followed by the Nysa family (8.2\%). The comparable probabilities obtained for Flora and Polana indicate that the dynamical information alone does not provide a unique identification of the parent family.

Although S-type asteroids dominate the Flora family, it shares orbital-element space with the Baptistina family, whose members are commonly associated with the X complex \citep{Nesvorny2026}. 
The X-type asteroids typically possess moderate geometric albedos of $p_V \sim 0.1$ \citep{Mahlke2022}, aligning with the value measured for \WN. 
Consequently, an origin within the dynamically overlapping Flora--Baptistina complex cannot be ruled out based on its albedo and reflectance spectrum.

The Nysa-Polana complex is also compositionally heterogeneous. While its higher-albedo component primarily consists of S-type asteroids linked to Nysa, the primitive Polana-Eulalia component displays a continuum of B- and C-type spectra \citep{Cellino2001,2016Icar..266...57D}. Because \WN exhibits a B-type-like spectrum, its compositionally aligns best with the Polana-Eulalia population, pointing toward a primitive carbonaceous origin. Nevertheless, the overlapping dynamical probabilities prevent an unambiguous assignment to a single parent family.
Further spectroscopic observations are required to better constrain the composition of \WN and determine its origin.

\subsection{Implications for the Close Approach in June 2028} \label{subsec:res_ca}
\WN is known to make an extremely close approach in 2028 \citep{Dotson2022}. 
It is an exceptional opportunity to image the disk of such a small target, 
without requiring a rendezvous space mission.
To plan disk-resolved imaging observations, 
several quantities are key parameters.
The projected angular diameter is obviously the most critical parameter
but both the apparent rate and
apparent magnitude will be limiting factors
(owing to the large non-sidereal motion of \WN, the adaptive optics
loop must be closed on the target itself, which thus needs to be bright enough).

Based on our analysis, which supports the classification of \WN as a B-type object, we derived its physical properties by calculating the absolute magnitude in the $V$ band, $H_V$, using the observed $g$ and $r$ magnitudes \citep{Tonry2012}.
The resulting values are 
$H_V = 18.31 \pm 0.06$ mag for the $H$-$G$ model and $H_V = 18.50 \pm 0.05$ mag for the linear model. 
By combining these results with the derived geometric albedo, the effective diameter was calculated. 
The $H$-$G$ model yielded a diameter of $0.81\pm0.13$~km, while the linear model resulted in $0.74\pm0.11$~km. 
This diameter agrees with the reported diameters of
$0.85$, 
$0.932_{-0.011}^{+0.011}$, and
$0.94 _{-0.39 }^{+0.39 }$\,km
based on NEATM analysis of 
NEOWISE mid-infrared photometry by
\citet{2022PSJ.....3...30M},
\citet{2011ApJ...743..156M} and
\citet{2021PSJ.....2..162M}.
For the purpose of this study,
we adopted the intermediate value of 0.78 km as the representative diameter for our calculations, 
since the presence or absence of an opposition surge remains uncertain.

\begin{figure}[ht]
\centering
\includegraphics[width=1.0\hsize]{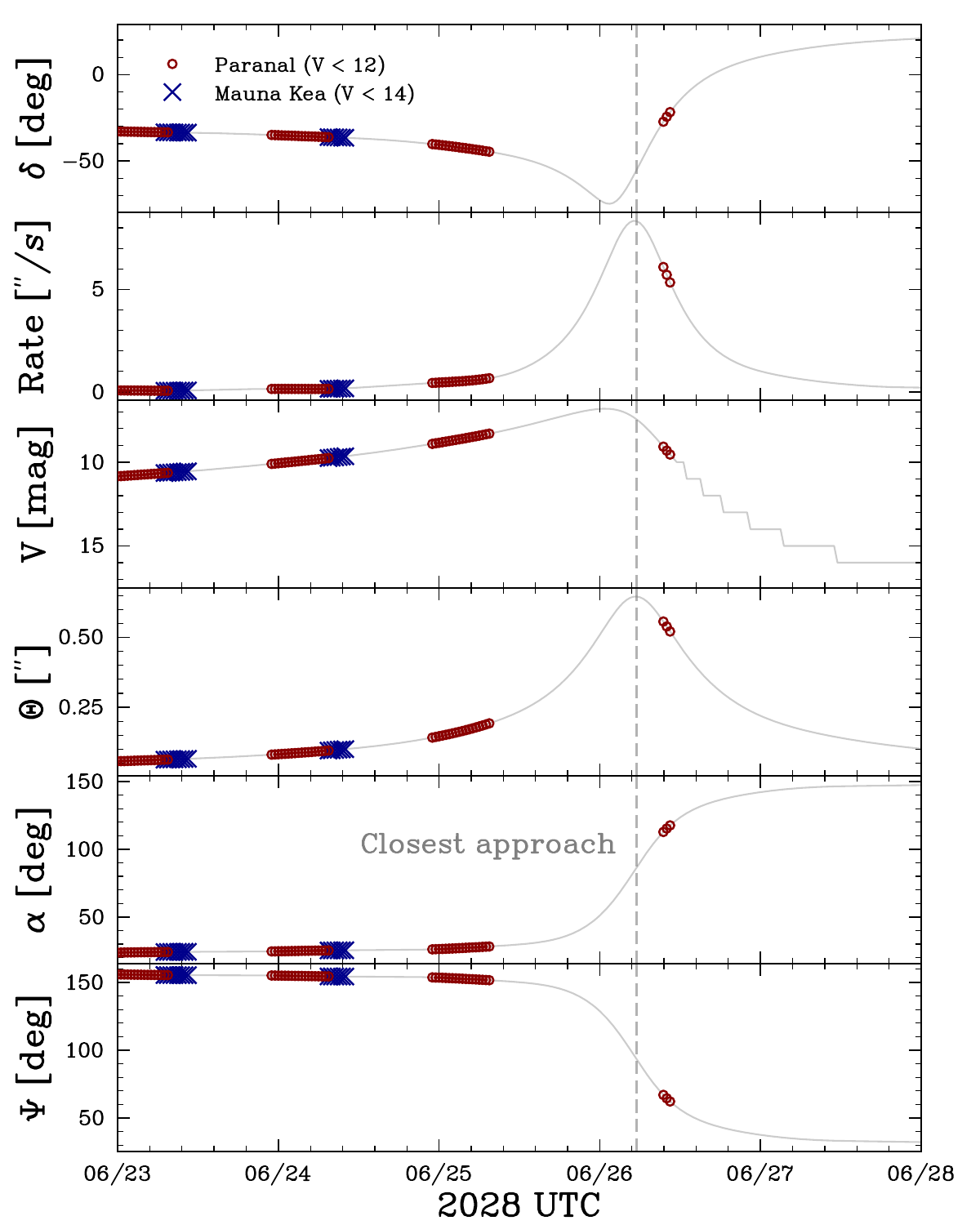}
\caption{
    Predicted
    declination ($\delta$),
    rate,
    $V$ magnitude,
    angular size ($\Theta$),
    phase angle ($\alpha$), and
    solar elongation ($\Psi$)
    of \WN around its closest approach
    on June 26, 2028.
    Expected windows of observations (see text) from
    Paranal (VLT) and Mauna Kea (Keck/Subaru) are marked.
    }
\label{fig:size}
\end{figure}

We present in Fig.~\ref{fig:size} the ephemerides of \WN around its closest approach,
obtained with 
the NASA JPL Horizons system 
using the Python package \texttt{astroquery} \citep{Ginsburg2019}.
We checked the expected observability of \WN from all largest facilities equipped with
adaptive-optics (AO) cameras:
ESO VLT/SPHERE,
Keck/NIRC2,
LBTO/SHARK-VIS,
GTC/FRIDA,
Subaru/SCExAO.
We set limit of airmass below two,
the Sun at least 12\degr below the horizon (nautical twilight), 
a solar elongation of at least 40\degr, 
and an apparent V mag brighter than 12 for ESO/SPHERE and 14 for the other facilities.

\WN is not observable from LBTO or GTC.
It can be imaged for two nights before the
closest approach from both ESO VLT in Chile, and Hawaiian telescopes (Keck and Subaru).
On 24-06-2028,
\WN will reach an angular diameter of about 0.1\arcsec,
providing limited details on its shape but allowing to probe its vicinity 
for satellites \citep[e.g.,][]{2022Icar..38215013V, 2025A&A...698A.136M},
a unique opportunity to measure its mass \citep{2012PSS...73...98C}.
On June 25, 2028, its angular diameter will reach 0.2\arcsec, allowing surface features
to be finely imaged. It represents a rare opportunity to resolve a NEA, such angular diameter being more typical of 100-200\,km diameter 
main-belt asteroids \citep[see, e.g.][]{
2017A&A...604A..64M,
2021A&A...650A.129C,
2021A&A...654A..56V}.

Provided VLT/SPHERE can track and keep the AO loop closed
at a non-sidereal rate of 5--7 \arcsec/s which is challenging,
on June 26, 2028, \WN will have an angular diameter larger than 0.5\arcsec, 
that only asteroids like Pallas and Ceres attain
\citep{2008A&A...478..235C, 2020NatAs...4..569M}.
This is significantly larger than the apparent diameter of 
NEAs previously
imaged from ground-based AO facilities
(308635)~2005~YU$_{55}$ with Keck-II/NIRC2 \citep[0.12\arcsec,][]{Merline2012} and (66391)~Moshup~(1999~KW$_4$) with VLT/SPHERE \citep[component separation of $\sim0.06$\arcsec,][]{Reddy2022}. It opens a unique possibility to detail the surface of a sub-kilometric
asteroid, something generally only achieved with a space mission.
Furthermore, inhomogeneous surface properties of NEAs have been reported in previous studies \citep[e.g.,][]{Lopez-Oquendo2022}, and we can directly investigate the surface heterogeneity of \WN using spatially resolved images in different filters and with integral field spectroscopy
\citep[such as with SPHERE/IFS,][]{2008-SPIE-Claudi, 2018Icar..309..134P}.

A comprehensive characterization of the shape and surface properties of \WN is still lacking, which is notable given that it will be one of the most accessible large NEAs. 
The next observing window for disk-integrated photometry
spans from late 2026 to early 2027, during which \WN is expected to be brighter than $V$\,$\sim$\,22 from early November 2026 to mid-February 2027. 
As shown in Fig.~\ref{fig:loc}, the observing geometry during this apparition differs significantly from previous ones, making this an important opportunity to better constrain the lightcurve, sidereal period, and shape as well as to enable a more reliable interpretation of the 2028 close approach from ground-based observations.

\section{Conclusion}
In this work, we have presented a comprehensive physical characterization of the potentially hazardous asteroid (153814)~\WN by consolidating new photometry from DK154 with extensive archival data from TESS, DECam, ZTF, and ATLAS. 
This multi-epoch, multi-instrument dataset enabled us to refine the synodic rotation period to $4.263 \pm 0.002$ h from DK154 and $4.259 \pm 0.009$ h from TESS.
The observed lightcurve amplitudes exhibit a large variation ranging from 0.12 to 0.63~mag. 
The complex lightcurve morphology probably reflects a non-ellipsoidal shape.
Although severe degeneracies remain that prevent a unique determination of the sidereal rotation period, our spin-pole searches favor a preferred pole orientation located in the southern hemisphere in ecliptic coordinates.

Four independent color datasets from ATLAS, DECam, ZTF, and DK154 consistently place \WN within the C-complex, resolving previous taxonomic inconsistencies in the literature. 
By integrating these colors with published near-infrared spectra and our phase curve constraints, which yield a geometric albedo of $p_{\rm V} = 0.13 \pm 0.04$ from the TESS data, the object is most accurately classified as a B-type asteroid under the Mahlke taxonomy. 
This taxonomic assignment yields a refined effective diameter of 
$D = 0.81 \pm 0.13$~km using the $H$-$G$ model, or
$D = 0.74 \pm 0.11$~km based on the linear phase model,
both of which are in excellent agreement with independent thermal-infrared estimates.
Dynamical modeling further suggests that \WN most likely originated in the inner main belt through the $\nu_6$ resonance, with the Flora--Baptistina and Nysa--Polana complexes emerging as the plausible parent candidates.

The upcoming encounter on June 26, 2028, at a geocentric distance of approximately 0.65~lunar distances, offers a premier observational opportunity to investigate a kilometer-scale carbonaceous NEA at high angular resolution. 
During its closest approach, the asteroid is expected to reach a maximum apparent angular diameter of about 0.5~arcsec, making it an ideal target for ground-based disk-resolved imaging using adaptive optics systems (such as VLT/SPHERE) as well as planetary radar facilities. 
To fully capitalize on this rare event, targeted photometric and lightcurve observations during the 2026–2027 apparitions will be essential to collect data at complementary viewing geometries, break the current period-pole degeneracies, and mature the 3D shape model before the close flyby.

\begin{acknowledgements}
We thank  
Robert D. Stephens, Brian D. Warner,
and Brian A. Skiff for providing the original photometric data through the ALCDEF database.
We would like to thank the anonymous referees for their valuable comments, which have greatly improved the manuscript.
This work was supported by JSPS KAKENHI Grant Number 25H00665.
This work was supported by the French government through the France 2030
investment plan managed by the National Research Agency (ANR), as part of the
Initiative of Excellence Université Côte d’Azur under reference number ANR-15-IDEX-01.
This work was supported by the Japan Society for the Promotion of Science (JSPS) Overseas Research Fellowships.
The work at Ond\v{r}ejov was supported by the {\it Praemium Academiae} award (no. AP2401) from the Academy of Sciences of the Czech Republic.
This research used the facilities of the Canadian Astronomy Data Centre operated by the National Research Council of Canada with the support of the Canadian Space Agency.
This work uses data obtained from the Asteroid Lightcurve Data Exchange Format (ALCDEF) database, which is supported by funding from NASA grant 80NSSC18K0851.
This work uses data from the University of Hawaii's ATLAS project, funded through NASA grants NN12AR55G, 80NSSC18K0284, and 80NSSC18K1575, with contributions from the Queen's University Belfast, STScI, the South African Astronomical Observatory, and the Millennium Institute of Astrophysics, Chile.
This paper includes data collected with the TESS mission, obtained from the MAST data archive at the Space Telescope Science Institute (STScI). Funding for the TESS mission is provided by the NASA Explorer Program. STScI is operated by the Association of Universities for Research in Astronomy, Inc., under NASA contract NAS 5–26555.
The Legacy Surveys consist of three individual and complementary projects: the Dark Energy Camera Legacy Survey (DECaLS; Proposal ID \#2014B-0404; PIs: David Schlegel and Arjun Dey), the Beijing-Arizona Sky Survey (BASS; NOAO Prop. ID \#2015A-0801; PIs: Zhou Xu and Xiaohui Fan), and the Mayall z-band Legacy Survey (MzLS; Prop. ID \#2016A-0453; PI: Arjun Dey). DECaLS, BASS and MzLS together include data obtained, respectively, at the Blanco telescope, Cerro Tololo Inter-American Observatory, NSF’s NOIRLab; the Bok telescope, Steward Observatory, University of Arizona; and the Mayall telescope, Kitt Peak National Observatory, NOIRLab. Pipeline processing and analyses of the data were supported by NOIRLab and the Lawrence Berkeley National Laboratory (LBNL). The Legacy Surveys project is honored to be permitted to conduct astronomical research on Iolkam Du’ag (Kitt Peak), a mountain with particular significance to the Tohono O’odham Nation.
NOIRLab is operated by the Association of Universities for Research in Astronomy (AURA) under a cooperative agreement with the National Science Foundation. LBNL is managed by the Regents of the University of California under contract to the U.S. Department of Energy.
This project used data obtained with the Dark Energy Camera (DECam), which was constructed by the Dark Energy Survey (DES) collaboration. Funding for the DES Projects has been provided by the U.S. Department of Energy, the U.S. National Science Foundation, the Ministry of Science and Education of Spain, the Science and Technology Facilities Council of the United Kingdom, the Higher Education Funding Council for England, the National Center for Supercomputing Applications at the University of Illinois at Urbana-Champaign, the Kavli Institute of Cosmological Physics at the University of Chicago, Center for Cosmology and Astro-Particle Physics at the Ohio State University, the Mitchell Institute for Fundamental Physics and Astronomy at Texas A\&M University, Financiadora de Estudos e Projetos, Fundacao Carlos Chagas Filho de Amparo, Financiadora de Estudos e Projetos, Fundacao Carlos Chagas Filho de Amparo a Pesquisa do Estado do Rio de Janeiro, Conselho Nacional de Desenvolvimento Cientifico e Tecnologico and the Ministerio da Ciencia, Tecnologia e Inovacao, the Deutsche Forschungsgemeinschaft and the Collaborating Institutions in the Dark Energy Survey. The Collaborating Institutions are Argonne National Laboratory, the University of California at Santa Cruz, the University of Cambridge, Centro de Investigaciones Energeticas, Medioambientales y Tecnologicas-Madrid, the University of Chicago, University College London, the DES-Brazil Consortium, the University of Edinburgh, the Eidgenossische Technische Hochschule (ETH) Zurich, Fermi National Accelerator Laboratory, the University of Illinois at Urbana-Champaign, the Institut de Ciencies de l’Espai (IEEC/CSIC), the Institut de Fisica d’Altes Energies, Lawrence Berkeley National Laboratory, the Ludwig Maximilians Universitat Munchen and the associated Excellence Cluster Universe, the University of Michigan, NSF’s NOIRLab, the University of Nottingham, the Ohio State University, the University of Pennsylvania, the University of Portsmouth, SLAC National Accelerator Laboratory, Stanford University, the University of Sussex, and Texas A\&M University.
BASS is a key project of the Telescope Access Program (TAP), which has been funded by the National Astronomical Observatories of China, the Chinese Academy of Sciences (the Strategic Priority Research Program “The Emergence of Cosmological Structures” Grant \# XDB09000000), and the Special Fund for Astronomy from the Ministry of Finance. The BASS is also supported by the External Cooperation Program of Chinese Academy of Sciences (Grant \# 114A11KYSB20160057), and Chinese National Natural Science Foundation (Grant \# 12120101003, \# 11433005).
The Legacy Survey team makes use of data products from the Near-Earth Object Wide-field Infrared Survey Explorer (NEOWISE), which is a project of the Jet Propulsion Laboratory/California Institute of Technology. NEOWISE is funded by the National Aeronautics and Space Administration.
The Legacy Surveys imaging of the DESI footprint is supported by the Director, Office of Science, Office of High Energy Physics of the U.S. Department of Energy under Contract No. DE-AC02-05CH1123, by the National Energy Research Scientific Computing Center, a DOE Office of Science User Facility under the same contract; and by the U.S. National Science Foundation, Division of Astronomical Sciences under Contract No. AST-0950945 to NOAO.
ZTF is supported by the National Science Foundation under Grants No. AST-1440341 and AST-2034437 and a collaboration including current partners Caltech, IPAC, the Oskar Klein Center at Stockholm University, the University of Maryland, University of California, Berkeley , the University of Wisconsin at Milwaukee, University of Warwick, Ruhr University, Cornell University, Northwestern University and Drexel University. Operations are conducted by COO, IPAC, and UW.
This work was developed within the Fink community and made use of the Fink
resources. Fink is a broker for LSST
\citep{Moller2021} and is supported by LSST-France and CNRS/IN2P3.
The properties of \WN were retrieved thanks to the \texttt{SsODNet} service
\citep{2023A&A...671A.151B}
of the Space Service (SE-OP)
of Laboratoire Temps Espace at Paris Observatory through
its Solar System Portal\footnote{\url{https://ssp.imcce.fr}}
and \texttt{rocks} python interface\footnote{\url{https://rocks.readthedocs.io/en/latest/}}.
\end{acknowledgements}

\bibliographystyle{aa} 
\input {aa61418-26.bbl}

\begin{appendix}
\nolinenumbers
\section{Photometry of DECam images} \label{app:cutout} 
\begin{figure*}[ht]
\centering

\begin{subfigure}[b]{0.22\hsize}
  \includegraphics[width=\linewidth]{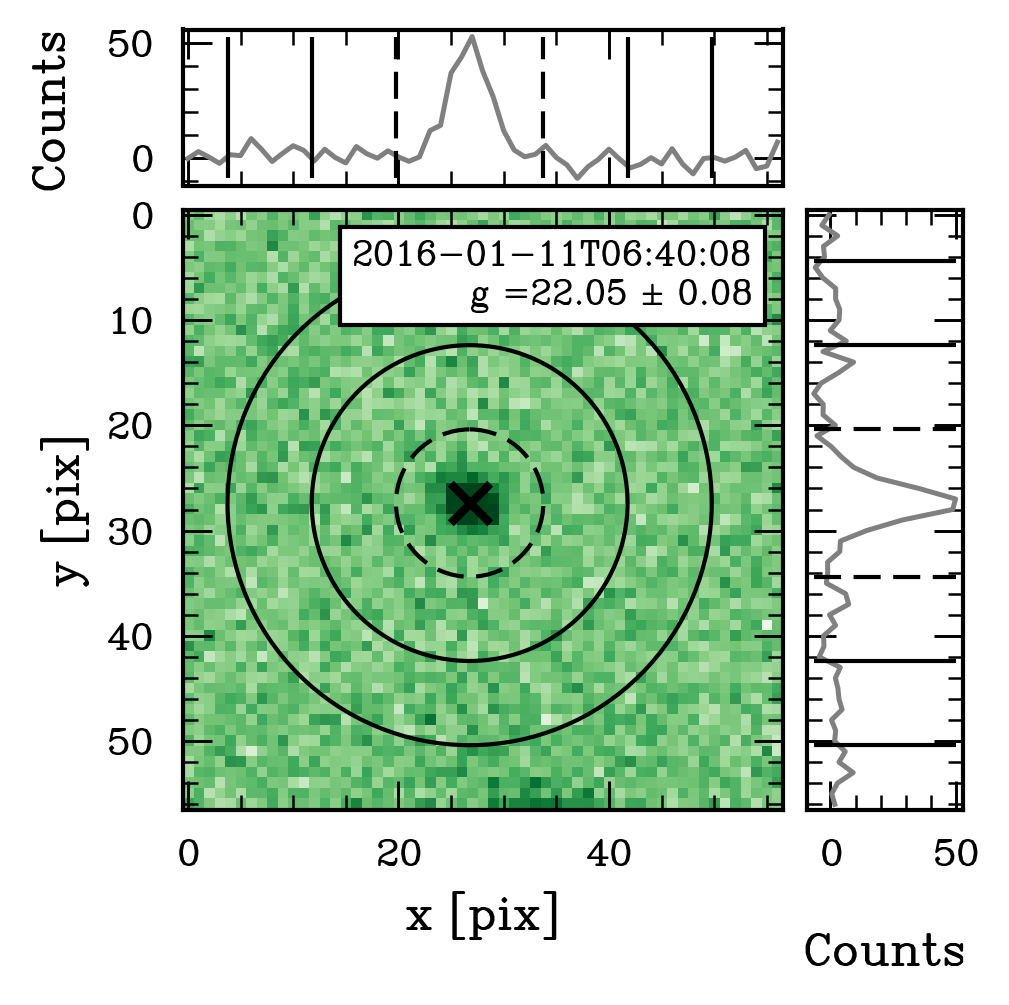}
\end{subfigure}
\hfill 
\begin{subfigure}[b]{0.22\hsize}
  \includegraphics[width=\linewidth]{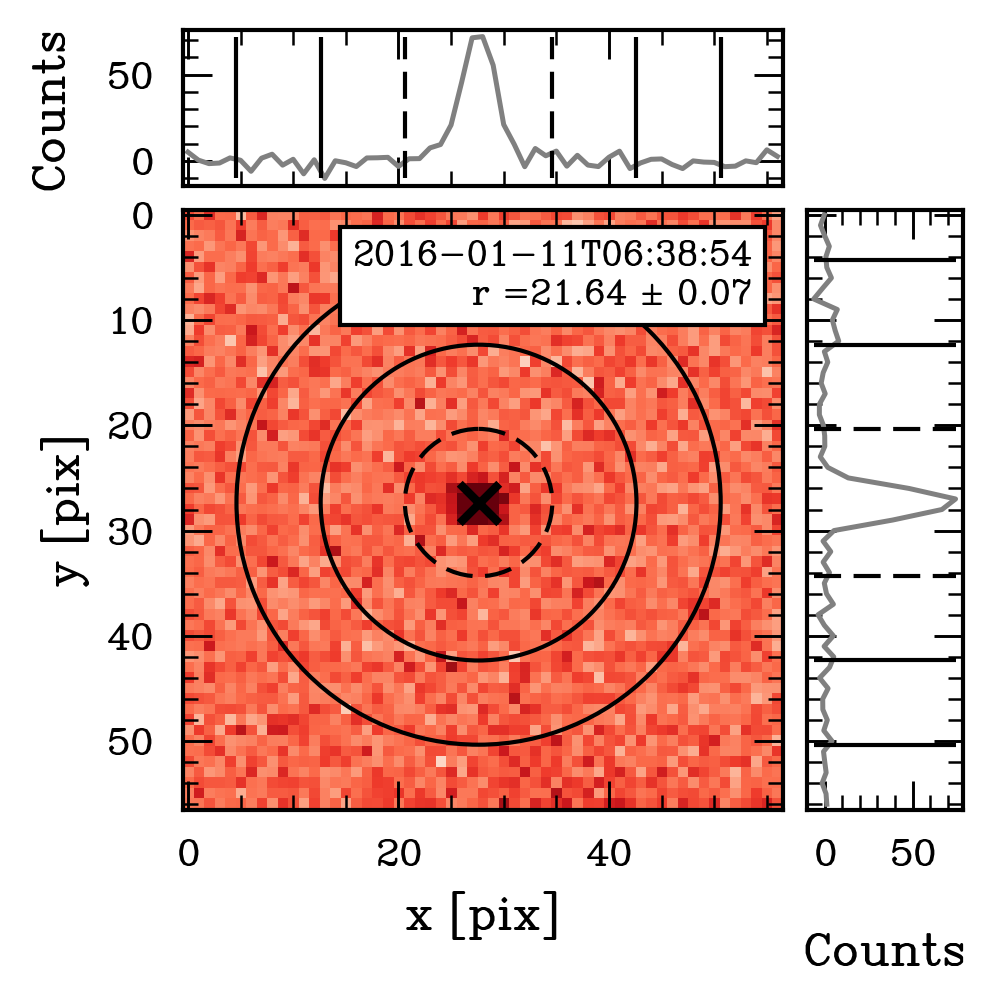}
\end{subfigure}
\hfill
\begin{subfigure}[b]{0.22\hsize}
  \includegraphics[width=\linewidth]{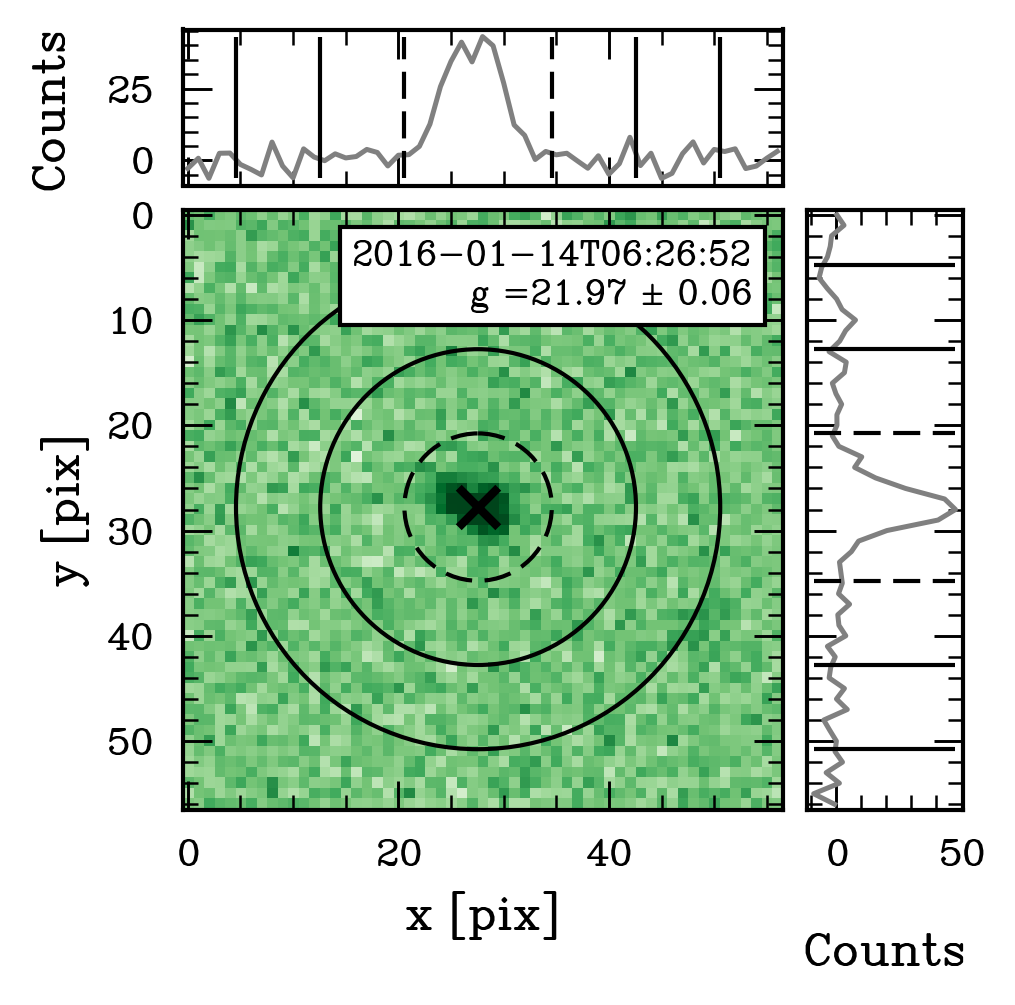}
\end{subfigure}
\hfill
\begin{subfigure}[b]{0.22\hsize}
  \includegraphics[width=\linewidth]{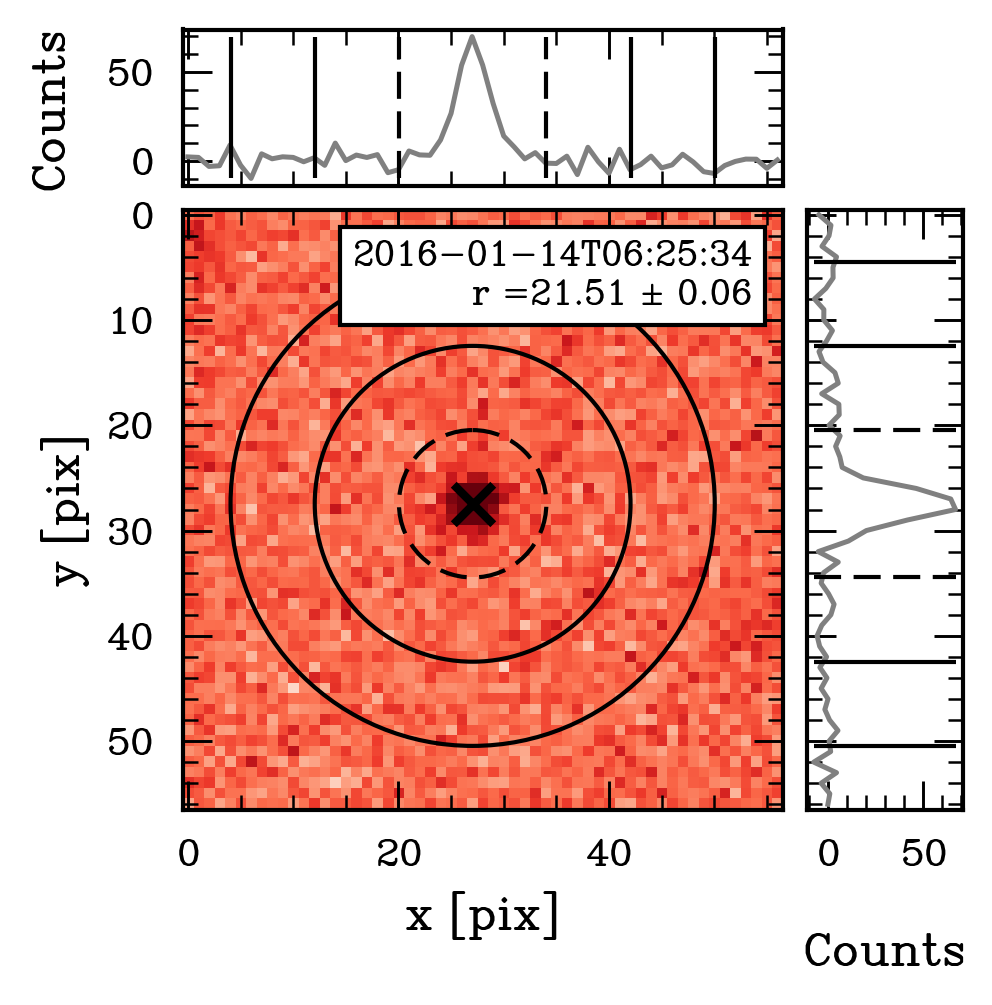}
\end{subfigure}

\vspace{1mm} 

\begin{subfigure}[b]{0.22\hsize}
  \includegraphics[width=\linewidth]{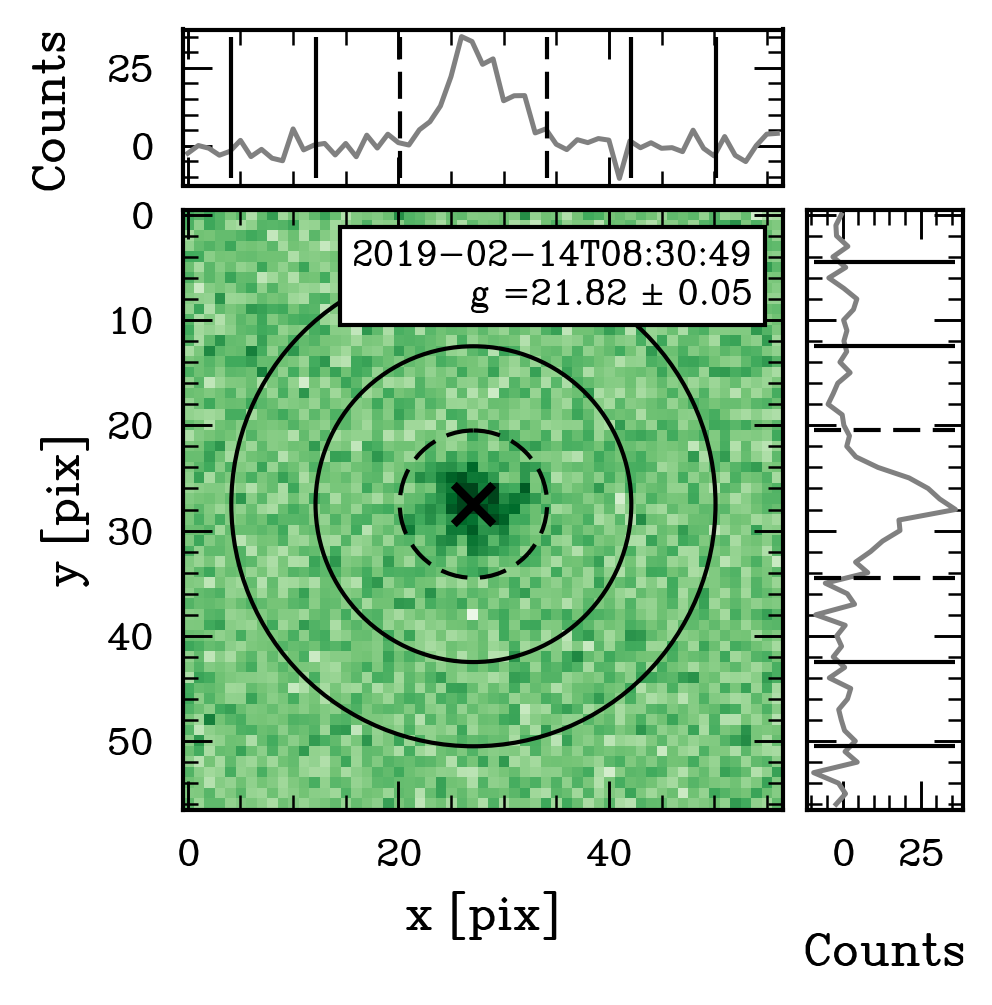} 
\end{subfigure}
\hfill
\begin{subfigure}[b]{0.22\hsize}
  \includegraphics[width=\linewidth]{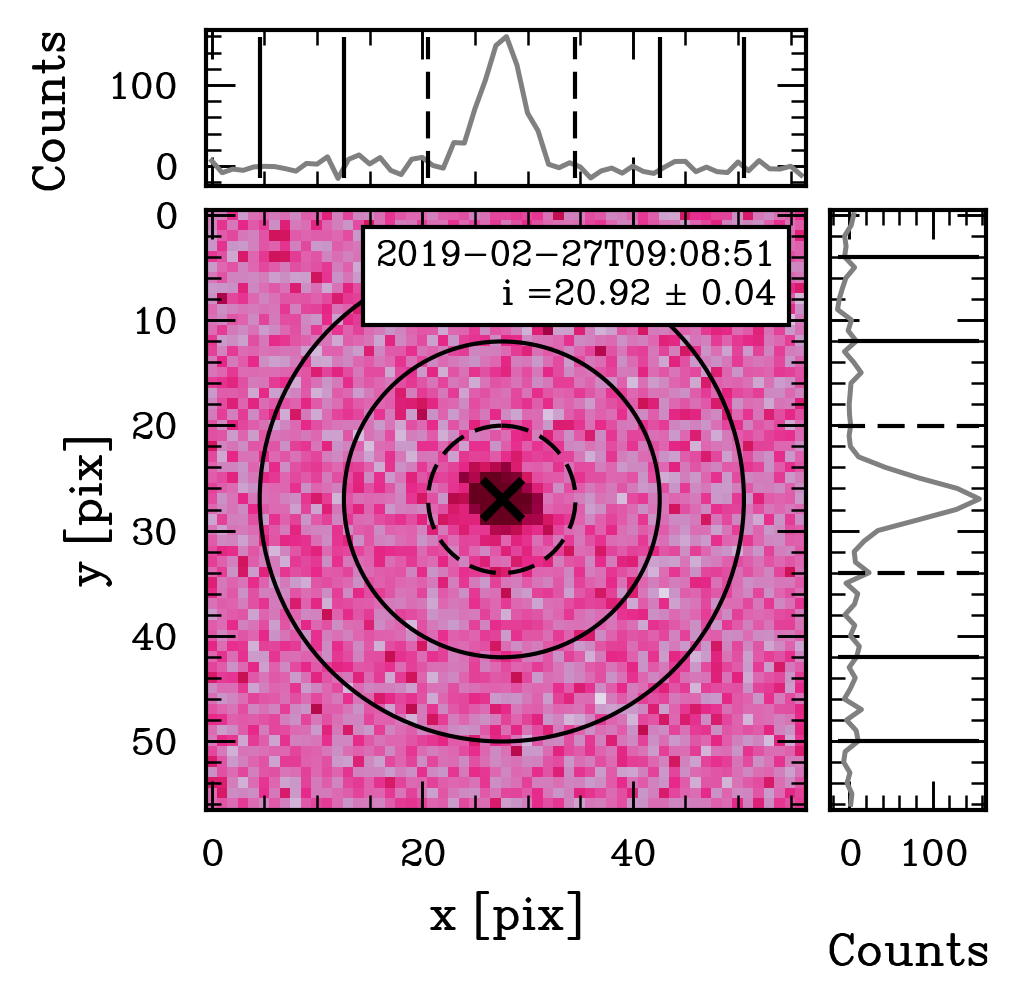}
\end{subfigure}
\hfill
\begin{subfigure}[b]{0.22\hsize}
  \includegraphics[width=\linewidth]{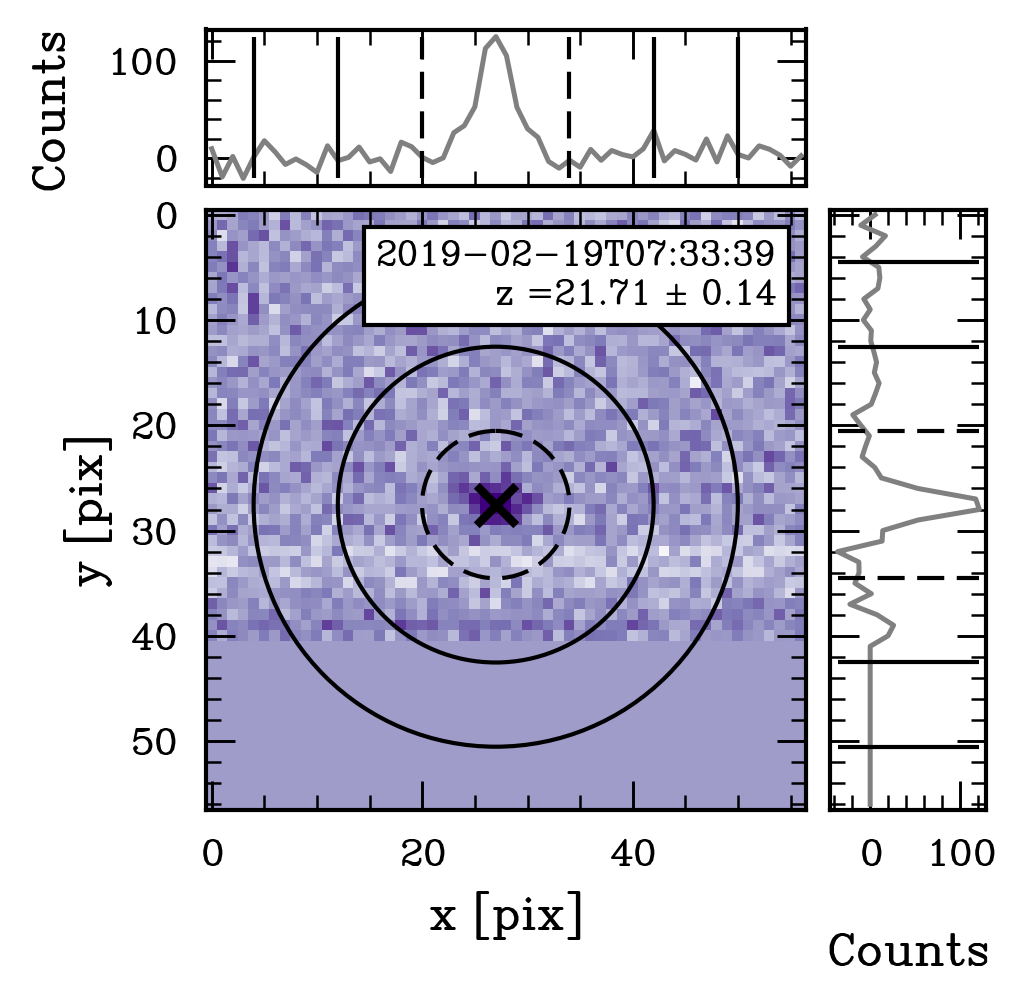}
\end{subfigure}
\hfill
\begin{subfigure}[b]{0.22\hsize}
  \includegraphics[width=\linewidth]{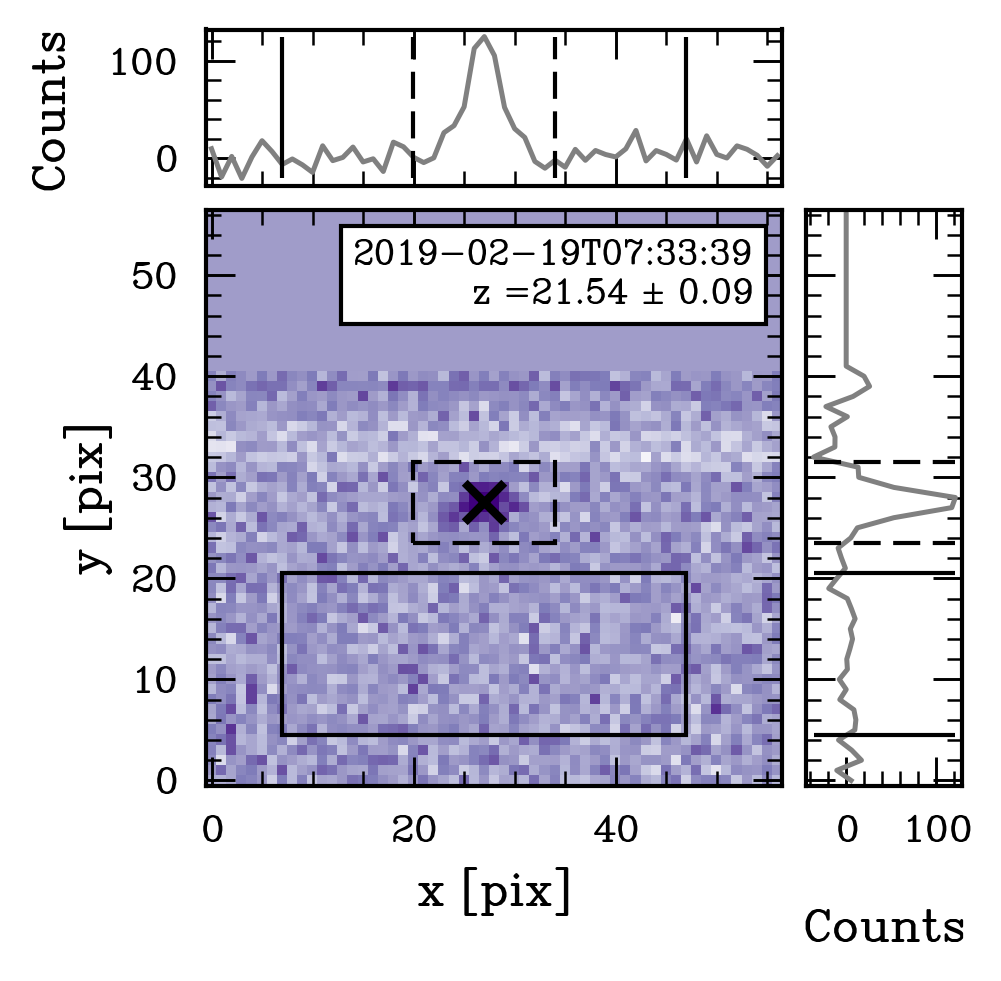}
\end{subfigure}

\caption{
    DECam detections of \WN. 
    Crosses indicate \WN.
    Field of view covers 15~arcsec$\times$15~arcsec.
 North is to the top and East is to the left.
 The dashed region represents the aperture used for photometry, and the region enclosed by solid lines indicates the area used for local background estimation.
}
\label{fig:cutout}
\end{figure*}

The resulting cutout images are presented in Fig. \ref{fig:cutout}. 
In these panels, dashed lines indicate the apertures used for photometry, 
while solid lines denote the annuli used for background and noise estimation.
For the $z$-band specifically, the rectangular aperture necessitated by the chip-edge artifacts is also plotted.

\section{Shape modeling} \label{app:shape}
\subsection{Lightcurve inversion}
First, we used the lightcurve inversion method developed in \citet{Kaasalainen2001a}, 
implemented in the public code provided by the Database of Asteroid Models from Inversion Techniques \citep[DAMIT;][]{Durech2010}, following the methodology of \citet{Fatka2025}. 
We used both dense lightcurves and ATLAS sparse photometry 
for the modeling, with the exception of TESS observations, which were excluded due to their low signal-to-noise ratio and long effective exposure times relative to the target’s rotation period. 
Following the standard convention, we first determined the sidereal rotation period using \texttt{period\_scan}.

\begin{figure}[ht]
\centering
\includegraphics[width=1.0\hsize]{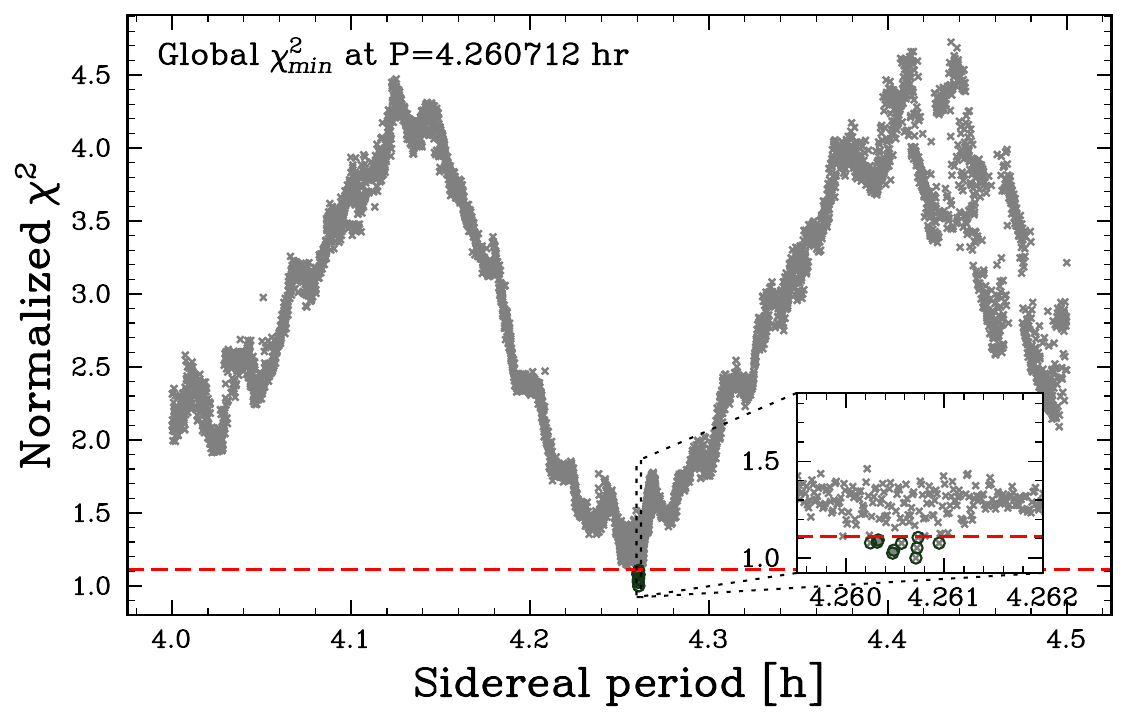}
\caption{
    Results of sidereal period search with \texttt{period\_scan}.
    The dashed horizontal line indicates the 3$\sigma$ threshold. 
    Circles indicate solutions below the 3$\sigma$ threshold.
    The region around the global minimum is shown in the inset.
    }
\label{fig:ps_wATLAS}
\end{figure}

We estimate 3$\sigma$ uncertainties of parameters with the following boundary:
\begin{equation}
\chi^2 \leq \chi^2_{\mathrm{min}}(1 + 3\sqrt{2/\nu}), \label{eq:bound_V17}
\end{equation}
where 
$\chi^2_{\mathrm{min}}$ is the global minimum chi-squared 
and $\nu$ is the degrees of freedom.
In this analysis, the number of data points is 1462 and the number of parameters is 17, yielding a total degrees of freedom of $\nu = 1445$.
Eq. (\ref{eq:bound_V17}) is used in, e.g., \citet{Vokrouhlicky2017, Hanus2018},
which is an approximate expression, 
but as the resulting minimum chi-squared is close to unity, they yield nearly identical results. 
While in some cases the period can be well constrained even from two dense lightcurves separated by approximately a decade \citep{Fatka2025}, in the present case this approach produced multiple degenerate period solutions (Fig.\ref{fig:ps_wATLAS}). 
We could not constrain a unique sidereal period.

Subsequently, we constrained the spin-axis orientation by searching for the global chi-squared minimum in the pole parameter space. 
Since a unique sidereal rotation period could not be determined due to multiple comparable minima,
we adopted the global minimum period as a reference solution, and derived a corresponding global best-fit model, 
including the spin axis and shape.
We set the sidereal period as a free parameter and explored the corresponding shape and spin-axis solutions.
To search for the pole direction of \WN, we generated 2001 pole orientations using the golden spiral algorithm.
In this process, only the spin-axis orientation is fixed,
and sidereal period and shape are optimized.
In this analysis, the number of parameters is 49, yielding a total degrees of freedom of $\nu = 1413$.

\begin{figure}
    \includegraphics[width=1.0\hsize]{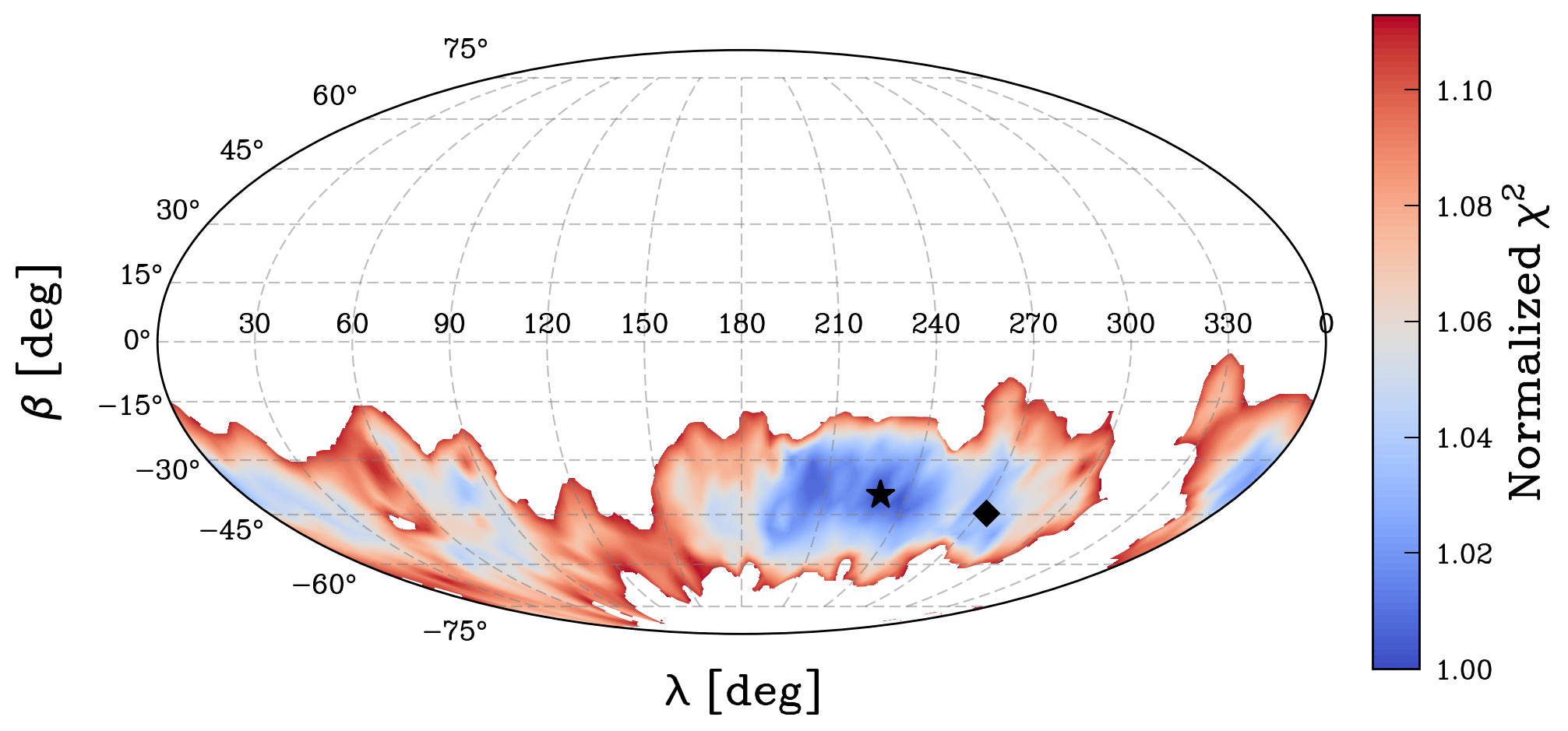}
\caption{
Results of the pole search. Only solutions below the 3$\sigma$ threshold are shown. The global minimum at $(\lambda, \beta) = (229.9^\circ, -39.0^\circ)$ is marked with a star.
The alternative solution used for the shape reconstruction is marked with a diamond at $(\lambda, \beta) = (272.9^\circ, -44.5^\circ)$.
}
\label{fig:pole}
\end{figure}

\begin{figure}[ht]
\centering
\includegraphics[width=0.9\hsize]{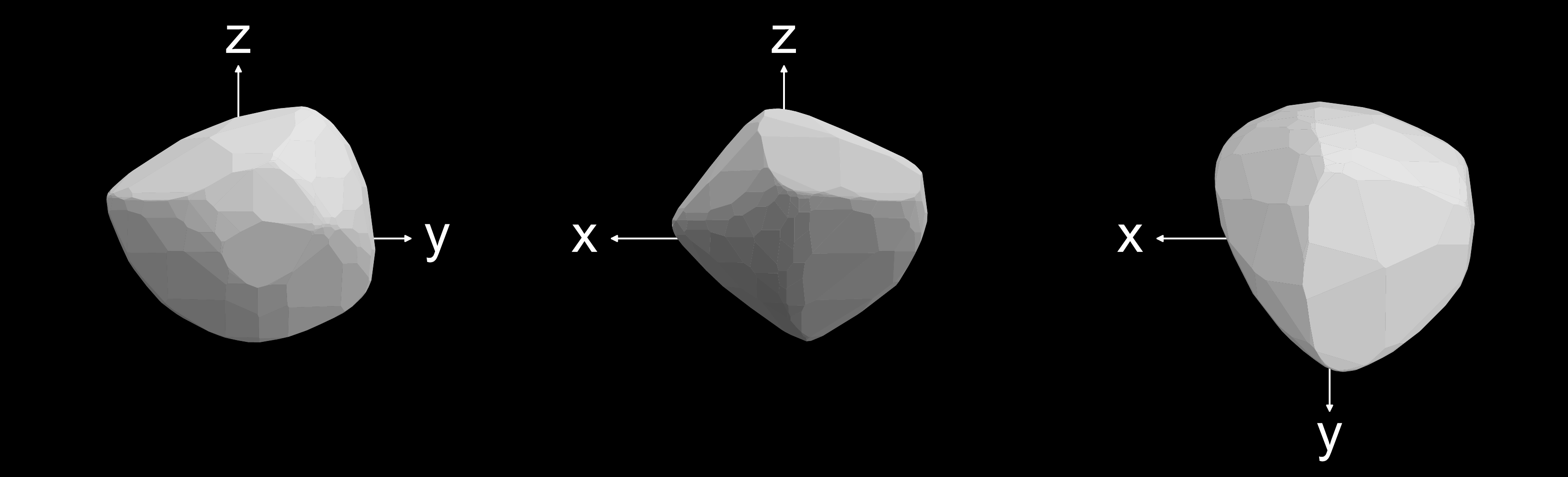}
\caption{
    Shape model reconstructed the lightcurve inversion method.
    The equatorial view is shown in the left and center panels, while the pole-on view is displayed on the right.
    }
\label{fig:shape_damit}
\end{figure}

The result of pole search is shown in Fig.~\ref{fig:pole}. 
The best fit values are $(\lambda, \beta) = (229.9^\circ, -39.0^\circ)$ (star in Fig.~\ref{fig:pole}). 
The corresponding shape model was unstable and physically unrealistic because its longest axis was the z-axis (rotation axis). 
Therefore, we filtered out this unrealistic solution \citep{Kaasalainen2001a} and estimated the shape using the alternative solution, $(\lambda, \beta) = (272.9^\circ, -44.5^\circ)$ (diamond in Fig.~\ref{fig:pole}), as shown in Fig.~\ref{fig:shape_damit} as a representative example.
This analysis indicates that the preferred pole solutions are located in the southern hemisphere in the ecliptic coordinate. 
The SOCCA solution discussed in the next subsection is also within the $3\sigma$ threshold.

\subsection{SOCCA}
The SOCCA phase curve model \citep{xenos26}
was also applied to the dataset. The model corrects for the viewing geometry effects by fitting a triaxial ellipsoid to the photometric observations, simultaneously retrieving the ellipsoid shape parameters, spin axis orientation and sidereal roatational period.

The best-fit solution yielded an RMS residual of 0.12 mag. The relatively large residual is expected as the triaxial ellipsoid approximation cannot fully reproduce the irregular shape features of the object. The model recovers the sidereal rotation period of $4.260727 \pm 8.4 \times 10^{-6}$ hours. The spin-axis orientation was determined to be $(\alpha_0,\delta_0) = (62.42^\circ, -29.36^\circ) \pm (1.98^\circ, 3.6^\circ)$, corresponding to ecliptic coordinates $(\lambda,\beta) = (51.85^\circ, -49.21^\circ)$. 
The spin axis orientation and rotation period given here by SOCCA correspond to the ellipsoid parameters providing the lowest fit RMS error and are therefore reported as the model's unique solution. While this pole differs from the convex shape model solution in ecliptic longitude by approximately $180^\circ$, the two solutions share similar $\beta$ values, reflecting a known mirror symmetry in pole determinations. 
The derived ellipsoidal shape parameters are $a/b = 1.18 \pm 0.02$ and $a/c = 2.09 \pm 0.39$ with $a>b>c$ the ellipsoid axes.
\end{appendix}

\end{document}

%% file: aa61418-26.bbl
\newcommand{\noopsort}[1]{} \newcommand{\singleletter}[1]{#1}